\documentclass{article}
\usepackage[utf8]{inputenc}
\usepackage[T1]{fontenc}
\usepackage{graphicx}
\usepackage{setspace}
\usepackage{xcolor}
\usepackage{subfigure}
\usepackage{amsmath,amssymb,amsfonts,graphicx}
\usepackage{verbatim,enumerate,hyperref,fancyhdr,setspace,array}
\usepackage[margin=3cm]{geometry}
\usepackage{authblk}
\usepackage{multirow}
\usepackage{bm}
\usepackage{natbib}
\usepackage{wrapfig}
\usepackage{hyperref}

\newcommand{\blind}{0}      

\begin{document}

\if0\blind
{
\title{Coping with Information Loss and the Use of Auxiliary Sources of Data: A Report from the NISS Ingram Olkin Forum Series on Unplanned Clinical Trial Disruptions}
\author[1,$\dagger$]{Silvia Calderazzo}
\author[2,$\dagger$]{Sergey Tarima}
\author[1,]{Carissa Reid}
\author[3]{Nancy Flournoy}
\author[4,5]{Tim Friede}
\author[6]{Nancy Geller}
\author[7]{James L Rosenberger}
\author[8]{Nigel Stallard}
\author[9,10,11]{Moreno Ursino}
\author[12]{Marc Vandemeulebroecke}
\author[13]{Kelly Van Lancker}
\author[9,10]{Sarah Zohar\thanks{Corresponding author: \texttt{sarah.zohar@inserm.fr}}}

\affil[1]{Division of Biostatistics, 
DKFZ (German Cancer Research Center), Heidelberg, Germany}
\affil[2]{Division of Biostatistics, Medical College of Wisconsin, Wauwatosa, USA}
\affil[3]{University of Missouri, Columbia, Missouri, USA}
\affil[4]{Department of Medical Statistics, University Medical Center Göttingen, Göttingen, Germany}
\affil[5]{DZHK (German Center for Cardiovascular Research), partner site Göttingen, Göttingen, Germany}
\affil[6]{National Heart, Lung, and Blood Institute, National Institutes of Health, Bethesda, MD 20892-7913, USA}
\affil[7]{National Institute of Statistical Sciences, and Department of Statistics, Penn State University, University Park, PA 16802-2111 USA}
\affil[8]{Statistics and Epidemiology, Division of Health Sciences, Warwick Medical School, University of Warwick, Coventry, UK}
\affil[9]{Inserm, Centre de Recherche des Cordeliers, Université de Paris Cité, Sorbonne Université, Paris, France}
\affil[10]{Inria, HeKA, Inria Paris, France}
\affil[11]{Unit of Clinical Epidemiology, AP-HP, CHU Robert Debré, Université de Paris, Sorbonne Paris-Cité, Inserm CIC-EC 1426, Paris, France}
\affil[12]{Novartis Pharma AG, Basel, Switzerland}
\affil[13]{Department of Biostatistics, Johns Hopkins Bloomberg School of Public Health, Baltimore, USA}
\affil[$\dagger$]{Authors made equal contributions}
  \maketitle
} \fi

\if1\blind
{
  \bigskip
  \bigskip
  \bigskip
  \begin{center}
    {\LARGE\bf Coping with Information Loss and the Use of Auxiliary Sources of Data: A Report from the NISS Ingram Olkin Forum Series on Unplanned Clinical Trial Disruptions}
\end{center}
  \medskip
} \fi

\bigskip

\def\spacingset#1{\renewcommand{\baselinestretch}%
{#1}\small\normalsize} \spacingset{1}

\begin{abstract}
Clinical trials disruption has always represented a non negligible part of the ending of interventional studies. While the SARS-CoV-2 (COVID-19) pandemic has led to an impressive and unprecedented initiation of clinical research, it has also led to considerable disruption of clinical trials in other disease areas, with around 80\% of non-COVID-19 trials stopped or interrupted during the pandemic. In many cases the disrupted trials will not have the planned statistical power necessary to yield interpretable results. This paper describes methods to compensate for the information loss arising from trial disruptions by incorporating additional information available from  auxiliary data sources. The methods described include the use of auxiliary data on baseline and early outcome data available from the trial itself and frequentist and Bayesian approaches for the incorporation of information from external data sources. The methods are illustrated by application to the analysis of artificial data based on the Primary care pediatrics Learning Activity Nutrition (PLAN) study, a clinical trial assessing a diet and exercise intervention for overweight children, that was affected by the COVID-19 pandemic. We show how all of the methods proposed lead to an increase in precision relative to use of complete case data only.
\end{abstract}
\noindent

{\it{Keywords: interrupted studies, auxiliary data, external data, Bayesian inference, frequentist inference, statistical power}}

\newpage
\spacingset{1.45} 
\section{Introduction and Scope}

The notorious effect of clinical trial disruptions has been known for decades. Fogel et al. (2018) reported that failure in patients' recruitment adversely affected 25\% of cancer trials, 18\% of trials were closed with less than half of the target sample size, and 22\% of the failed phase 3 studies were closed due to lack of funding \cite{Fogel2018}. 
Reduction in power represents a major concern for clinical trials. 

For a while most of disrupted clinical trials were associated with "research waste" as the studies were stopped and data were rarely reused.   The SARS-CoV-19 (COVID-19) pandemic have exuberant this tendency as it had a disruptive effect on many ongoing clinical trials. While the number of registered clinical trials was not impacted by the pandemic \cite{ledford2020}, due to reorientation of medical research towards COVID-19, the number of completed clinical trials significantly dropped \cite{Hawila21}. Accordingly, it has been reported that around 80\% of non-COVID-19 trials were stopped or interrupted during the pandemic \cite{vanDorn2020}. Consequently, many disrupted studies no longer had the statistical power to yield interpretable results.

\begin{figure}\label{fig_scope}
\centering
\includegraphics[scale=0.5]{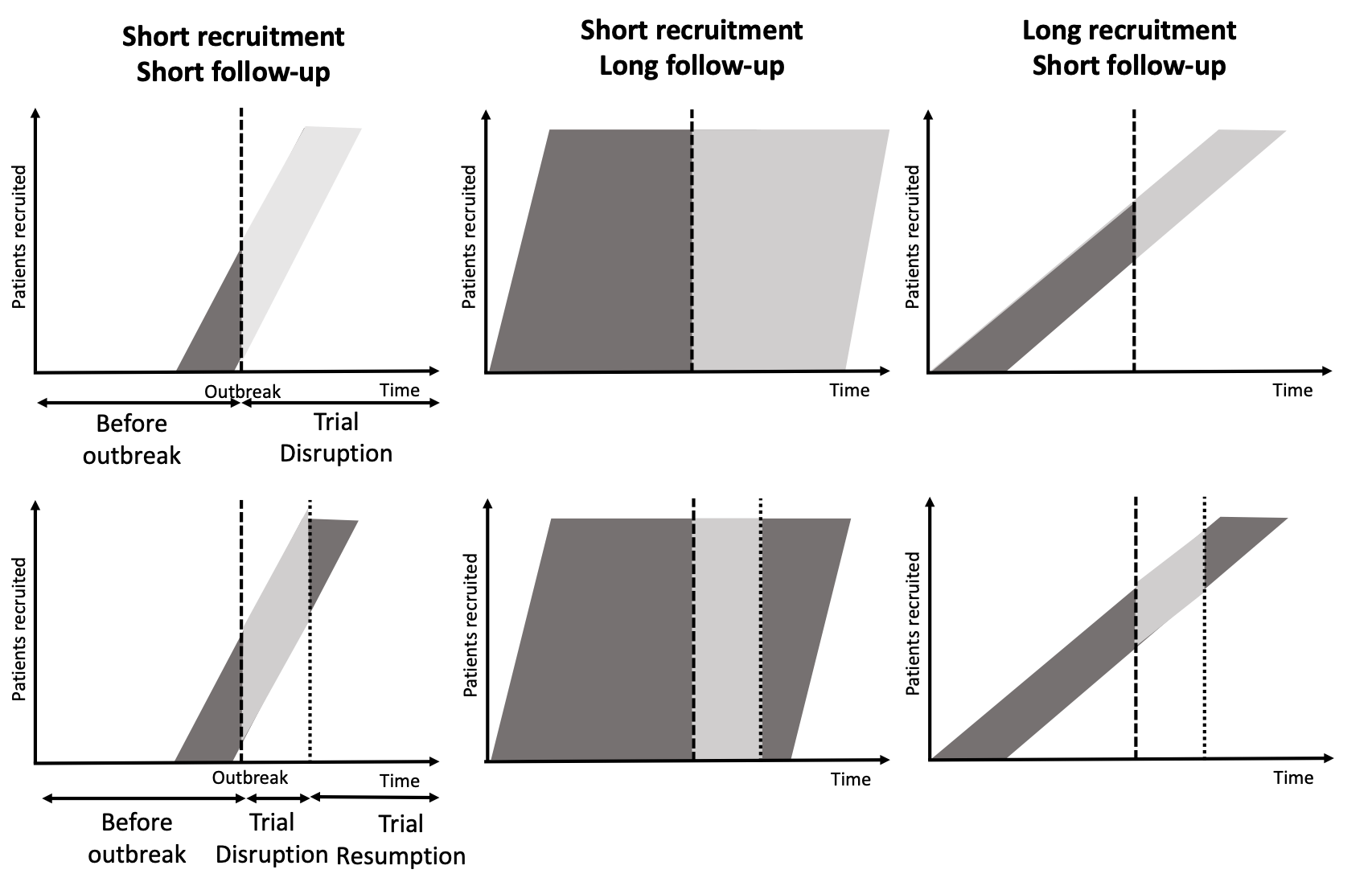}
\caption{Planned recruitment and duration of Clinical trials and the effect of information loss due to the interruption. This figure illustrates  the pandemic as an example but it can be the case in other interrupted trials. In each panel the dashed line indicates the time of onset of the pandemic and the dotted line indicates when (if at all) the trial resumed.  The dark grey areas to the left of the dashed line in all six panels (a-f) indicate the information obtained up to the start of the disruption.  The dark grey areas to the right of the dotted lines in panels (d,e,f)  illustrate that additional information was obtained once the trial resumed.  The light grey areas indicate the information loss due to the disruption.}
\end{figure}

Both recruitment and follow-up phases can be affected by clinical trials disruptions. Some studies can be temporarily paused and resumed later.   Others are halted permanently.  A few prevalent interruption patterns are shown in Figure \ref{fig_scope}.  Our manuscript deals primarily with studies closed at the interruption but some of the reported methods are applicable in more complex situations.

Depending on planned duration of recruitment and follow up, trials can be interrupted in somewhat different ways.  Figure \ref{fig_scope} illustrates the interruptions as a function of recruitment time and follow-up time.
Figure \ref{fig_scope}a  shows that even in a trial with short recruitment and short follow up, the power could be reduced substantially. Figure \ref{fig_scope}b  
 shows the impact of interruption on studies requiring long-term follow-up. This illustration ("short recruitment and long follow-up") , can be conveniently compared to the third case ("long recruitment and short follow-up") by graphing the interruption impact on studies with long recruitment phases..  Figures \ref{fig_scope}(a,b,c) \ref{fig_scope} illustrate interrupted trials stopped, by instance,  by the pandemic, whereas Figures \ref{fig_scope}(d,e,f) illustrate trials that were able to resume. 

The topic of {\it information loss} has been revitalized during the COVID-19 pandemic and a few methodological solutions were proposed in  \cite{akacha2020,kunz2020}. One solution is to compensate for the information loss by incorporating additional information available from  auxiliary data sources. These auxiliary data can be obtained  from one or several data sources, at the individual or aggregated levels, within the same experiment or from other trial(s). 

This paper  proposes and discusses a few statistical approaches for compensating for the information loss with auxiliary data in the context of interrupted and stopped Randomized Clinical Trials (RCT).

The paper is organized as follows. Section \ref{sec:statement} sets up technical notation and describes a clinical trial actually affected by COVID-19, the PLAN study. When writing the manuscript, the PLAN study was not yet completed and thus the trial data were unavailable. Therefore, artificial data were simulated imitating the inclusion criteria and overall settings of the PLAN study (see Section \ref{sec:casestudy}). Section \ref{sec:int} discusses the use of \textit{internal} auxiliary information such as early and baseline data that is available from the patients in the trial itself. Specifically, previously published methods  on adaptive designs with interim analyses efficiently use auxiliary data, which might be extracted from multiple sources (patients' care data, registries, clinical trials data, etc.). Section \ref{sec:ext} reviews a selection of frequentist and Bayesian methods for incorporating auxiliary information from external data sources. Focus is placed on assessing and addressing potential heterogeneity, both unforeseen and from known sources. Section \ref{sec:Discussion} discusses the results and their practical implications.

This manuscript is the product of a working group formed from ‘Session Five:  Coping with Information Loss and the Use of Auxiliary Sources of Data’ of the National Institute of Statistical Sciences (NISS) Ingram Olkin Forum Series on Unplanned Clinical Trial Disruptions which took place on March 23, 2021. For more information on this scholarly activity visit the event website: https://www.niss.org/events/ingram-olkin-forum-series-unplanned-clinical-trial-disruptions-0

\section{Statement of the problem}
\label{sec:statement}

\subsection{Notation} When a study is designed, researchers plan to enroll a sample of patients $\omega_1,\ldots,\omega_n$ from a population $\Omega$. The sample size $n$ could be determined from various considerations: desired operational characteristics, budgetary constraints or  estimation quality. When considering a two-armed RCT, each $i^{th}$ patient is randomized to a treatment or control regimen, defined by an  indicator $R_{i}=R(\omega_i)$. Let the primary study endpoint be denoted by $Y_i={Y}(\omega_i)$ and let $\mu_{r}=E(Y_i|R_i=r)$, $r=0,1$, where 0 denotes the control group and 1 denotes the treatment group. The main goal of the study is, then, to evaluate the treatment effect, which can be defined for example as $$\theta = \mu_1-\mu_0.$$ 
The endpoint may also depend on patient characteristics; the row vector $$\mathbf{X}_{i}=\mathbf{X}(\omega_i) = \left(X_1 (\omega_i),\ldots, X_p(\omega_i)\right),$$ 
contains the $p$ covariates of the $i$-th patient. In this manuscript, we follow the convention of writing vectors in bold and scalars in regular font. Let the probability distribution function of the study primary endpoint $Y_i$ conditional on the treatment assignment $R_i$  and patient covariates $\mathbf{X}_{i}$ be denoted by $f_{Y_i|R_i,\mathbf{X}_i}\left(Y_i |R_i,\mathbf{X}_{i},\boldsymbol\theta \right)$, where $\text{\boldmath$\theta$}=(\theta_1, \ldots, \theta_q)^{\prime}$ is a vector of parameters including the treatment effect $\theta$.
Let $Y_i$, $i=1,\dots,n$, be independent and identically distributed conditional on the treatment assignment and covariate values. Then 
the likelihood function of $\boldsymbol\theta$ is $L(\boldsymbol\theta; D_{(n)}) = \prod_{i=1}^n f_{Y_i|R_i,\mathbf{X}_i} \left( Y_i|R_i, \mathbf{X}_i, \boldsymbol\theta \right)$, where
the triplet $D_{(n)}=\{\mathbf{Y},\mathcal{X},\mathbf{R}\}$ aggregates all available data, i.e., $\mathbf{Y}=(Y_1,\ldots, Y_n)$,
$\mathbf{R}=(R_1, \ldots,R_n)$, and $\mathcal{X}$ is the $n \times p$ matrix of patient covariate values. Additionally, let $\hat{\boldsymbol\theta}_{(n)}$ denote the maximum likelihood estimator (MLE) based on $n$ observations, i.e., the value of $\text{\boldmath$\theta$}$ maximising $L(\text{\boldmath$\theta$}; D_{(n)})$.

When a surrogate endpoint or a short-term endpoint is available it will be denoted by ${Z}$ if needed. Then, $Z_i$  ($i=1, \dots, n$) 
has probability distribution function $f_{{Z_i|R_i,\mathbf{X}_i}}\left({Z_i}|R_i,\mathbf{X}_{i}, \boldsymbol\psi \right)$, where $\boldsymbol\psi$ is a vector of parameters $(\psi_1, \ldots, \psi_q)^{\prime}$.

Data from external sources may also be available to improve inferences about $\theta$. For this purpose, we introduce an additional subscript to the above notation, i.e., $R_{id},Y_{id},Z_{id},\mathbf{X}_{id},\boldsymbol{\theta}_d,\boldsymbol{\hat{\theta}}_{d(n_d)},D_{d(n_d)}$, where $D_{d(n_d)}=\{\mathbf{Y}_d,\mathcal{X}_d,\mathbf{R}_d\}$, and $d=0,\dots, D$ is the index for the data source. The current trial is identified by $d=0$, which is assumed when the subscript is omitted. 

\begin{figure}
\centering
\includegraphics[width=\linewidth]{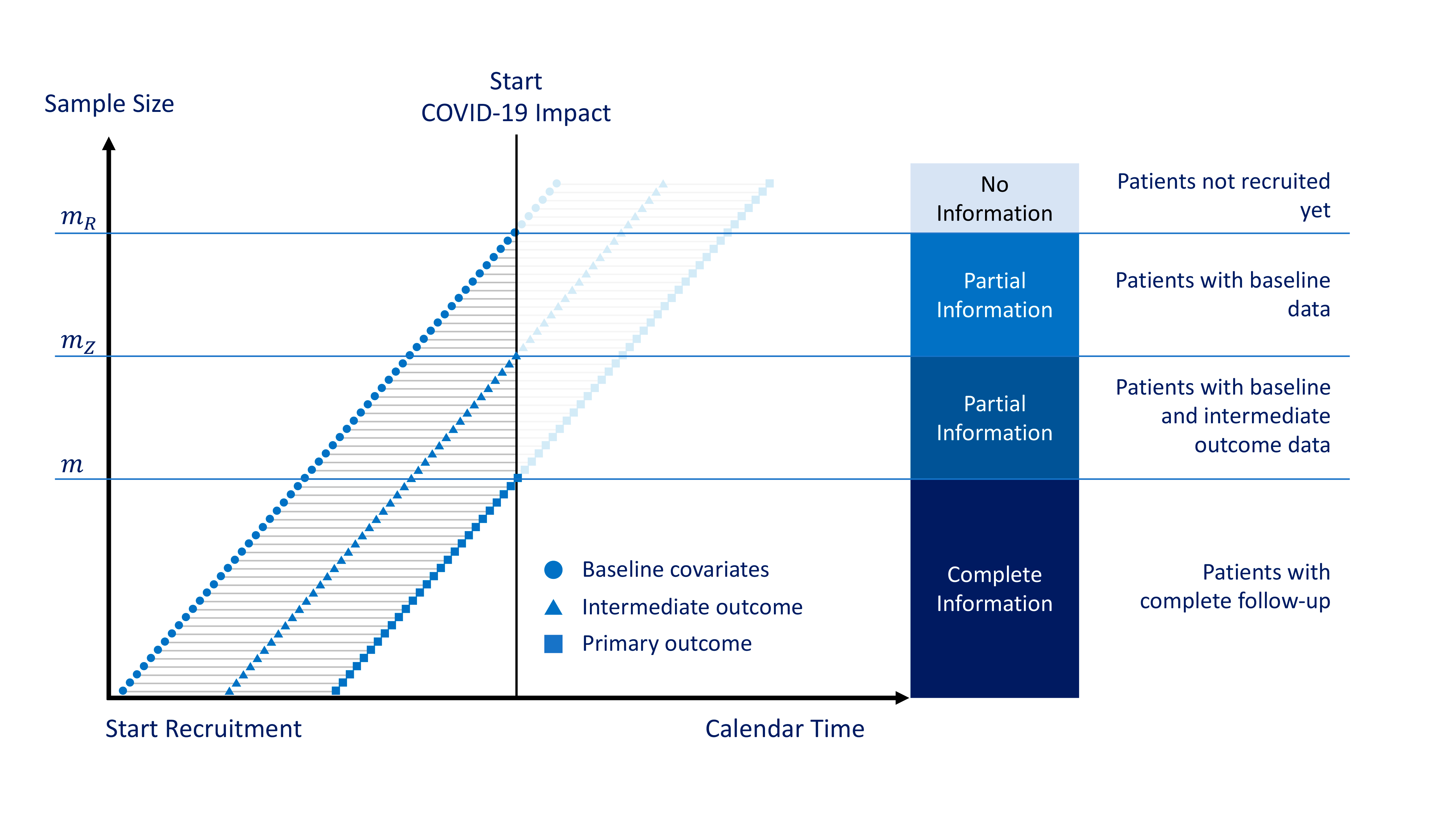}
\caption{Visualization of the available information at the start of the COVID-19 outbreak. The blue symbols show the available information, while the shadowed symbols show the unobserved information.}
\label{fig_impact}
\end{figure}

We assume that the first $m$ patients ($\omega_1,\ldots,\omega_m$) have completed the current study prior to the disruption. This means that $Y_1,\ldots,Y_m$ were unaffected by the disruption and $Y_{m+1},\ldots,Y_n$ may come from a different distribution.  Moreover, we assume that a short-term endpoint may have been additionally observed for $m_Z$ patients, and that baseline characteristics have been recorded for a potentially even larger number of recruited patients $m_R$, so that $m \leq m_Z \leq m_R \leq n$, as shown in Figure \ref{fig_impact}.

\subsection{Motivating Study and Artificial Illustrative Example}
\label{sec:casestudy}

Some trials that were well underway were able to be modified so that the original intent of the trial was maintained, but the intervention could be amended to adapt to the disruption.  For illustrative purposes we consider an artificial example based on the PLAN (Primary care pediatrics Learning Activity Nutrition) trial \cite{EPSTEIN2021106497}, a diet and exercise intervention for overweight children and one overweight parent compared to usual care. These dyads of an overweight child and parent were randomized to 
counseling (${R}_i=1$) or usual care (${R}_i=0$). Treatment was 26 or more counseling sessions over 24 months.  Per study protocol published at clinicaltrials.gov (NCT02873715), the plan was to enroll 528 dyads  with age and sex adjusted BMI percentile greater than $85\%$. The recruitment was completed with $452$ dyads ($n=452$). At the time of this writing, the PLAN trial was undergoing data analysis. 
\\ \indent
Thus, for illustrative purposes of this manuscript, the original study design is simplified and an artificial data set is generated as follows. The primary endpoint of the PLAN trial is change in \textit{age and sex adjusted body mass index percentile} (pBMI) of the child from baseline to 2 years; for the $i^{th}$ child the 2-year pBMI percentile change is $Y_{i}$. The short term endpoint is a 1-year pBMI percentile change ($Z_i$). In addition, another version of the study outcome is considered: a Z-score of the Pediatric BMI (zBMI). The planned analysis for the primary study outcome is the analysis of covariance, which compares intervention and control groups controlling for baseline BMI and  covariates. Thus, our illustrative example uses this planned statistical analysis. 
\\ \indent 
\textbf{Main dataset.} First, 452 left truncated Gaussian random variables with mean = 1.5 and unit variance are generated. The truncation region eliminates possibility of having observations below $1.036$ (85\% quantile of the standard normal distribution). These 452 random variables denoted as $zBMI1_i$ ($i=1,\ldots,452$) are baseline BMI values on the standardized Z-scale. The standard normal CDF ($\Phi(\cdot)$) applied to these 452 values generates baseline pBMI values [$pBMI1_i = \Phi(zBMI1_i)$] of 452 participants enrolled into the study. Second, the participants are randomized between the new intervention ($R=1$) and usual treatment ($R=0$) groups. For each study participant, a baseline covariate ($X$) is generated from $N(1,1)$. Third, 452 correlated pairs of Gaussian random variables $zBMI2_i$ and $zBMI3_i$ are generated from a bivariate normal with standard deviation equal to 0.3 and a correlation of 0.9. The mean structure is defined in Equations (\ref{EzBMI2}) and (\ref{EzBMI3}):
\begin{eqnarray}
E(zBMI2_i) &=& zBMI1_i -0.2 - 0.16 R_i  + 0.1 X_i \label{EzBMI2}\\
E(zBMI3_i) &=& zBMI1_i      - 0.08 R_i + 0.1 X_i.\label{EzBMI3}
\end{eqnarray}
Thus, the true treatment effect as assessed by the ANCOVA model is known and is equal to -0.08 on the Z-scale, and is approximately -0.01 on the percentile scale (pBMI).

\begin{figure}[ht!]
\centering
\subfigure[$pBMI,R=1$]{\includegraphics[width = 2.2in]{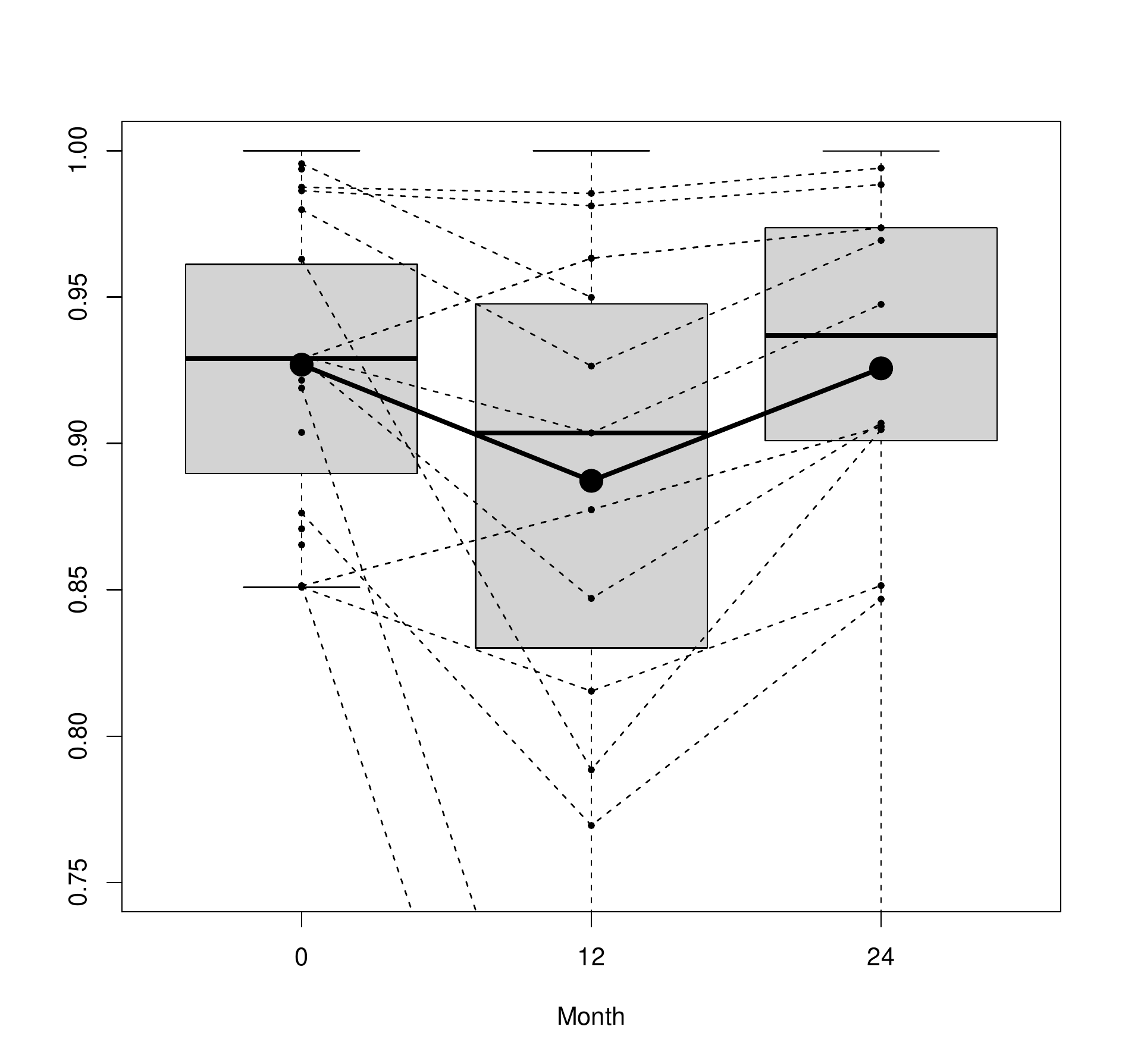}} 
\subfigure[$pBMI,R=0$]{\includegraphics[width = 2.2in]{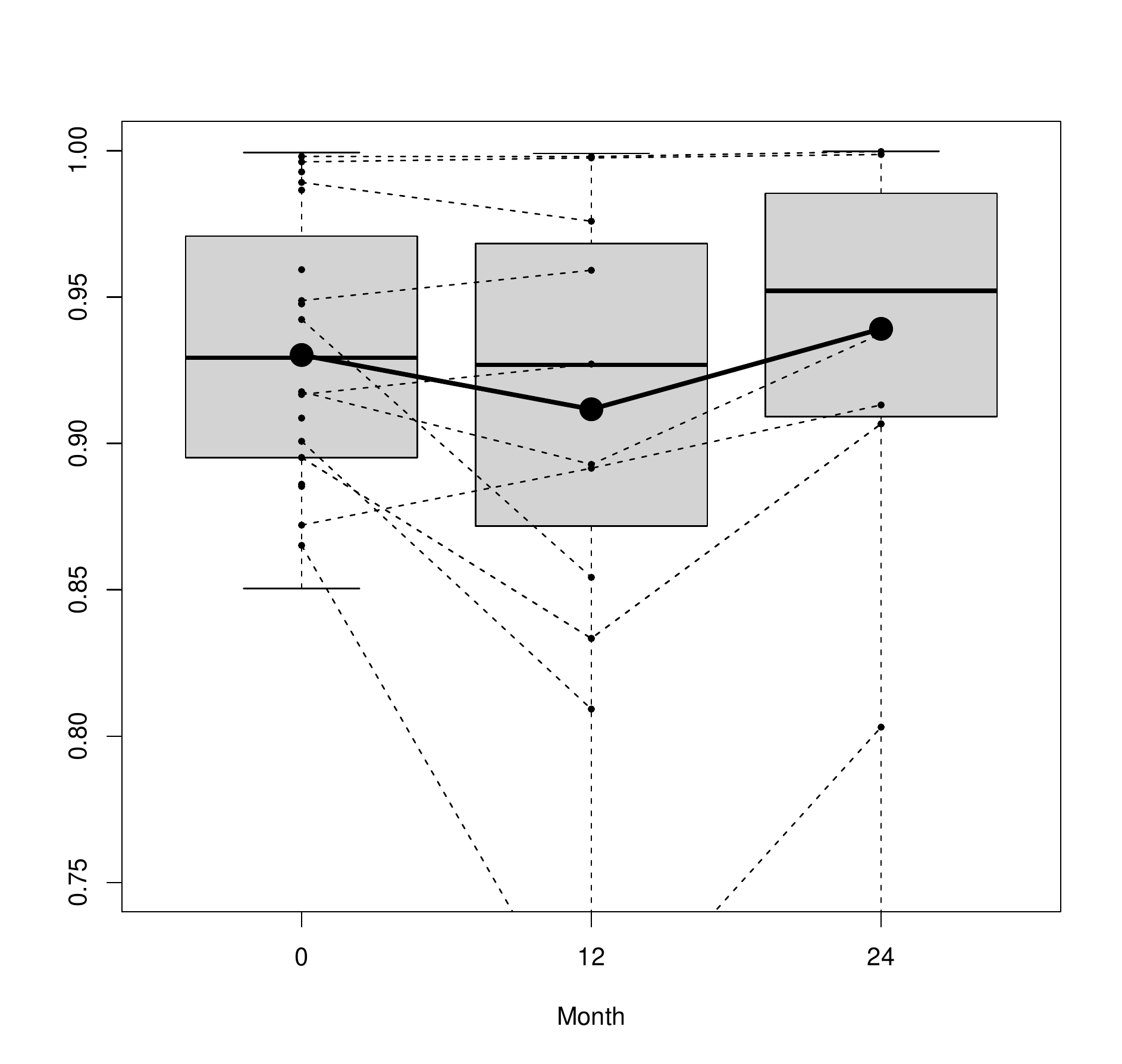}}
\subfigure[$zBMI,R=1$]{\includegraphics[width = 2.2in]{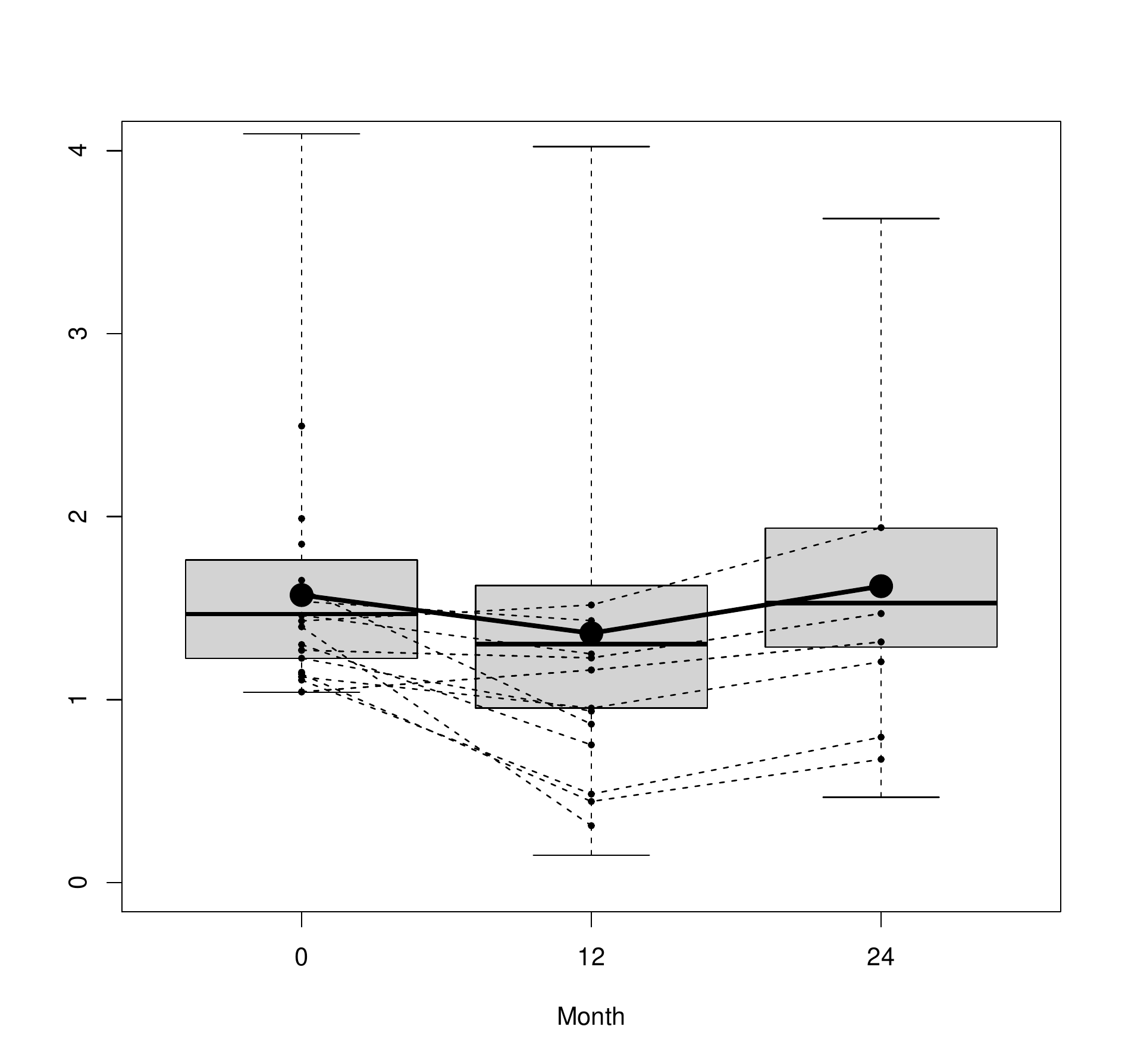}} 
\subfigure[$zBMI,R=0$]{\includegraphics[width = 2.2in]{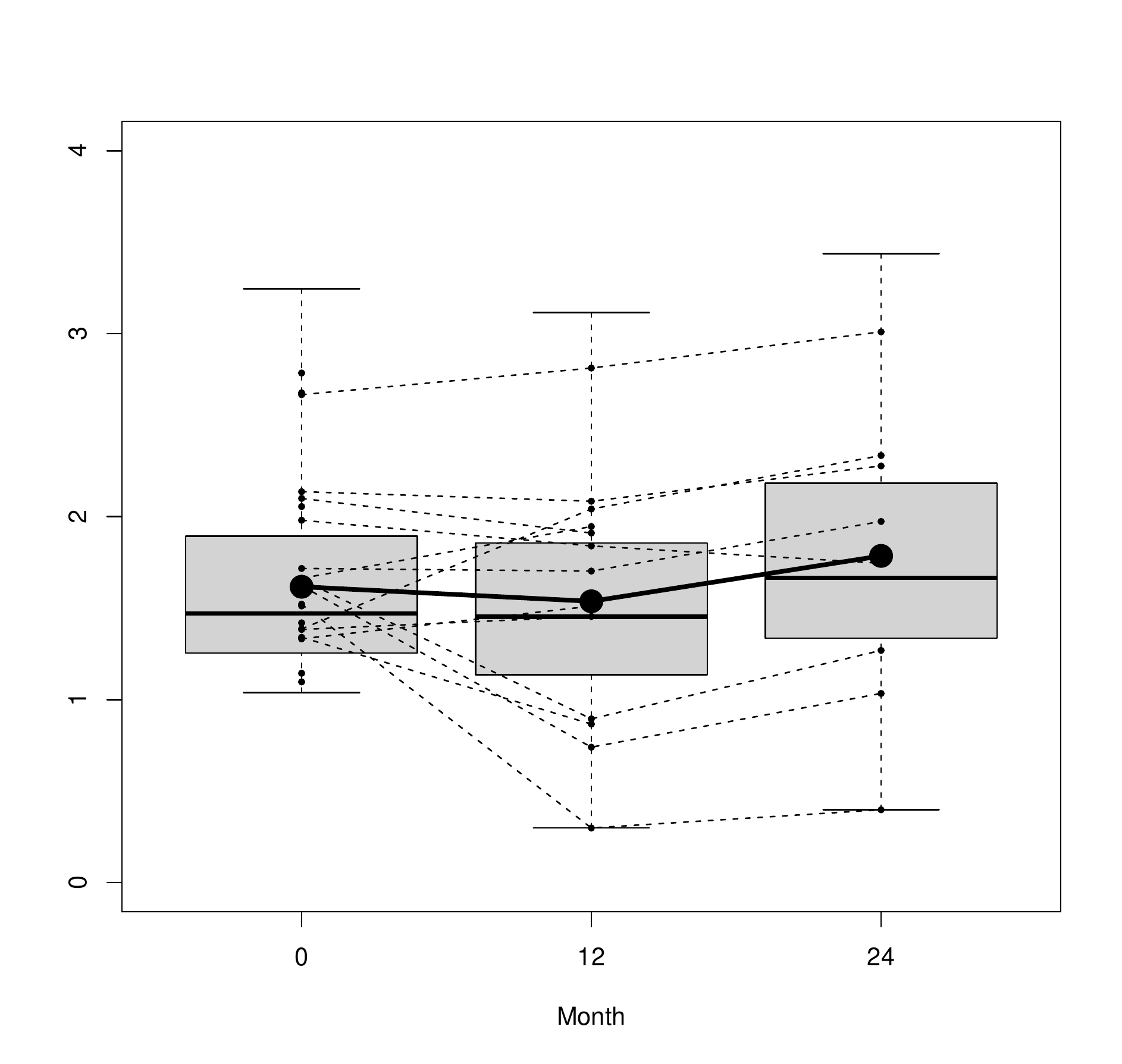}}
\caption{Boxplots, overtime profiles of 20 randomly selected participants and sample means summarizing the artificially created main dataset.
NOTE(NG):It is not obvious how this figure fits in with the text.  Someone needs to write a much more detailed caption and also to alter the text so that this figure fits in.}
\label{main_data}
\end{figure}

Fourth, $\Phi(\cdot)$ transforms  Gaussian data to the percentile scale: $pBMI2_i = \Phi(zBMI2_i)$ and $pBMI3_i = \Phi(zBMI2_i)$. Fifth,  125 randomly selected pairs of $pBMI2_i$ and $pBMI3_i$  are set to be missing. Finally, an additional 125 randomly selected $pBMI3_i$ from patients with observed $pBMI2_i$ are set to missing. Figure \ref{main_data} summarizes the generated dataset. 

In addition to the main dataset, we assumed that external data are available and generated artificial external datasets with a variety of sizes and treatment effects as follows: 
\begin{enumerate}
    \item[\textbf{(1)}] \textbf{External with the same sample size:} This dataset is generated from the same statistical model with the same sample size ($n=452$) as the main dataset, but it has no missing data.
    \item[\textbf{(2)}] \textbf{External with a larger sample size:} This dataset is generated from the same model as the main dataset, but its sample size ($n=904$) is twice as large and it has no missing data.
    \item[\textbf{(3)}] \textbf{External with a smaller sample size:} This dataset is generated from  the same model as the main dataset, but its sample size ($n=226$) is half as big and with no missing data.
    \item[\textbf{(4)}] \textbf{External with a different treatment effect:} This dataset is generated from  the same model as the main dataset,  with the same sample size  ($n=452$), but with no missing data and
    \begin{eqnarray}
    E(zBMI2_i) &=& zBMI1_i -0.2 - 0.36 R_i  + 0.1 X_i \label{EzBMI2_l}\\
    E(zBMI3_i) &=& zBMI1_i      - 0.28 R_i + 0.1 X_i\label{EzBMI3_l}.
    \end{eqnarray}
    \item[\textbf{(5)}] \textbf{External with a slightly different treatment effect:} This dataset is generated from  the same model as the main dataset,  with the same sample size  ($n=452$), but it has no missing data and
    \begin{eqnarray}
    E(zBMI2_i) &=& zBMI1_i -0.2 - 0.18 R_i  + 0.1 X_i \label{EzBMI2_s}\\
    E(zBMI3_i) &=& zBMI1_i      - 0.10 R_i + 0.1 X_i\label{EzBMI3_s}.
    \end{eqnarray}
\end{enumerate}
We repeated the analysis with  another version of the main dataset, named \textbf{main2}. In this new main dataset, we used a correlation of ($0.6$) between the 12-month and 24-month Pediatric BMI on the Z-scale.

\section{Internal Auxiliary Information}\label{sec:int}
\subsection{Sources of Auxiliary Information}

In this section we will consider the use of auxiliary information that is available from the patients in the trial itself.  In particular, we will describe methods for the use of early or baseline data in inference on the primary endpoint of interest.  

Using the notation introduced above, we envisage a study initially planned to include $n$ patients that is interrupted by the pandemic or any other reason so that primary endpoint data, $Y_i$, is available for patients $i=1, \ldots, m$ only, with $m <n$.  It is now desired to draw inference on the treatment effect $\theta$ based on these data.  The methods described below will use pre-interruption data only to avoid bias due to informative information loss, but will provide estimates with greater precision than those based on the observed $Y_1, \ldots, Y_m$ alone. 

Figure \ref{main_data} shows that the number of patients recruited to the trial prior to the disruption, which will be denoted $m_R$, will generally exceed $m$. In some cases, such as that of long follow-up and short recruitment, $m_R$ will be much larger than $m$. In many such settings, additional data on early endpoint(s) or baseline covariates will be available for all or some of patients $\omega_{m+1}, \ldots, \omega_R$ in addition to patients $\omega_1, \ldots \omega_m$.  

Suppose that, in addition to $Y_1, \ldots, Y_m$, we observe $\mathbf{X}_1, \ldots, \mathbf{X}_{m_R}$ and $Z_1, \ldots, Z_{m_Z}$ with $m \le m_Z \le m_R \le n$.
If $Z$ is a surrogate for $Y$, it may be of interest to use $Z_1, \ldots, Z_{m_Z}$ to estimate the treatment effect $\psi$ 
that is $E(Z \mid R=1) - E(Z \mid R=0)$. 
In other settings, the original treatment effect $\theta$ may remain of primary research interest.  In this case if $Y_i$ is correlated with $\mathbf{X}_i$ and/or $Z_i$ for $i=1, \ldots, m$, the additional information from $\mathbf{X}_{m+1}, \ldots, \mathbf{X}_{m_R}$ and/or $Z_{m+1}, \ldots, Z_{m_Z}$ may still be used as described in the following subsection. 

\subsection{Analytic Methods}

While the pandemic effect on clinical trials may be unprecedented, inference based on incomplete data as described above is not. For example, at an interim stage baseline and early endpoint data might be available while primary endpoint data is not.
  Analysis of data in such settings has been considered by a number of authors \cite{engel1991, marschner2001, galbraith2003, stallard2010, vanlancker2020}. 

\cite{engel1991} and \cite{galbraith2003} consider the case in which $Y_i$ and $Z_i$ have a bivariate normal distribution. Conditional on baseline covariates, their model is 
\begin{equation}
\label{eq:bivariatenormalcovariates}
\left (
\begin{array}{c} Y_i \\Z_i \end{array}
\right )
\sim N \left ( \left (
\begin{array}{c} a_Y+\mathbf{b}_Y ^\prime \mathbf{T}_i \\ a_Z+\mathbf{b}_Z ^\prime \mathbf{T}_i \end{array}
\right ),
\left (
\begin{array}{cc} \sigma_Y^2 & \rho \sigma_Y \sigma_Z \\ \rho \sigma_Y \sigma_Z & \sigma_Z^2 \end{array}
\right )
\right )
\end{equation}
where $\mathbf{T}_i=(\mathbf{X}_i,R_i)^{\prime}$.  The parameter of interest is the treatment effect on $Y_i$, that is $E(Y_i \mid R_i=1) - E(Y_i \mid R_i=0)$, which is given by the last element of $\mathbf{b}_Y$, and will be denoted $\theta$.

Model (\ref{eq:bivariatenormalcovariates}) can be rewritten as 
\begin{equation}
\label{eq:doubleregression1}
Z_i \sim N(a_Z + \mathbf{b}_Z^\prime \mathbf{T}_i, \sigma_Z^2) \\
\end{equation}
\begin{equation}
\label{eq:doubleregression2}
Y_i \mid Z_i \sim N(\alpha + \boldsymbol{\beta}^\prime \mathbf{T}_i + \gamma Z_i, \sigma_{Y|Z}^2)   
\end{equation}
with $\alpha = a_Y- \gamma a_Z$, $\boldsymbol{\beta} = \mathbf{b}_Y- \gamma \mathbf{b}_Z$, $\gamma= \rho \sigma_Y/\sigma_Z$ and $\sigma_{Y|Z}^2 = \sigma_Y^2 - \gamma^2 \sigma_Z^2 = \sigma_Y^2(1-\rho^2)$.  \cite{engel1991} call this a Double Regression model.

An estimate, $\hat {\mathbf{b}}_Z$, of $\mathbf{b}_Z$ can be obtained, together with an estimated variance, from a linear (least squares) regression model relating $Z_i$ to $\mathbf{T}_i$ for $i = 1, \ldots, m_Z$, while estimates, $\hat {\boldsymbol{\beta}}$ and $\hat \gamma$, of $\boldsymbol{\beta}$ and $\gamma$, respectively, can be obtained, together with their estimated variances and covariance, from a linear regression relating $Y_i, i = 1, \ldots, m$ to $\mathbf{T}_i$ and $Z_i$ for $i=1, \ldots, m$.  An estimate, $\tilde {\mathbf{b}}_Y$, of $\mathbf{b}_Y$ is then given by $\tilde {\mathbf{b}}_Y = \hat {\boldsymbol{\beta}} + \hat \gamma \hat {\mathbf{b}}_Z$. 

 \cite{engel1991} note that the variance of the estimate $\tilde {\mathbf{b}}_Y$ can be estimated by
\[
var(\hat {\boldsymbol{\beta}}) + \hat {\gamma}^2 var(\hat {\mathbf{b}}_Z) + 2 \hat {\mathbf{b}}_Z cov(\hat \beta, \hat \gamma) + \hat {\mathbf{b}}_Z^\prime \hat {\mathbf{b}}_Z var(\hat \gamma) 
\]
which may be obtained using the parameter and variance estimates from the linear regressions just described.

If $Z_i$ and $Y_i$ are correlated, a gain in precision  can be obtained by  using any additional early endpoint data $\lbrace Z_{m+1}, \ldots, Z_{m_Z}\rbrace$. \cite{galbraith2003} and \cite{stallard2010} show that in the simple case where there are no baseline covariates, so that $\mathbf{T}_i$ is a scalar indicator of treatment allocation and $\tilde {\mathbf{b}}_Y$ is the treatment effect estimate that we may write as $\tilde \theta$,  
\begin{equation}
\label{eq:marschnervariance}
var(\tilde \theta) = 
\left [ 1 - \rho^2 (1-\pi) \right ]
var(\hat \theta), 
\end{equation}
where $\hat \theta$ is the estimate of $\theta$ obtained directly from $\{Y_1, \ldots, Y_m\}$ alone 
and $\pi = m/m_Z$ is the proportion of patients with outcome $Z$ for whom outcome $Y$ is also observed.  The gain in precision using equation (\ref{eq:bivariatenormalcovariates}), with an additional covariate included, is illustrated by the example given in Section \ref{sec:application}.

A similar increase in precision  when  $Y_i$ and $Z_i$ have correlated Bernoulli distributions was observed by \cite{marschner2001}.
 For simplicity of notation, consider estimation in a single treatment group, where the parameter of interest is $p_Y = pr(Y_i=1)$.
Writing $p_Z = pr(Z_i=0)$, $p_{\bar Z} = pr(Z_i=0) = 1 - p_Z$, $p_{Y \mid Z} = pr(Y_i=1 \mid Z_i = 1)$ and $p_{Y \mid \bar Z} = pr(Y_i = 1 \mid Z_i=0) = (p_Y - p_Z p_{Y \mid Z})/p_{\bar Z}$, the maximum likelihood estimate of $p_Y$ is given by
\begin{equation}
\label{eq:marschnerestimate}
\tilde p_Y = \hat p_Z \hat p_{Y \mid Z} + \hat p_{\bar Z} \hat p_{Y \mid \bar Z}  
\end{equation}
where $\hat p_{Y \mid Z}$ and $\hat p_{Y \mid \bar Z}$ are estimated from $(Y_i, Z_i), i = 1, \ldots, m$ with $\hat p_{Y \mid Z} = \sum_{i=1}^m Y_i Z_i/\sum_{i=1}^m Z_i$ and $\hat p_{Y \mid \bar Z} = \sum_{i=1}^m Y_i (1-Z_i)/\sum_{i=1}^m (1-Z_i)$, and $\hat p_Z$ and $\hat p_{\bar Z}$ are estimated from $Z_i, i = 1, \ldots, m_Z$ with $\hat p_Z = \sum_{i=1}^{m_Z} Z_i/m_Z$ and $\hat p_{\bar Z} = \sum_{i=1}^{m_Z} (1-Z_i)/m_Z$.
Analogous to (\ref{eq:marschnervariance}),
\begin{equation}
\label{eq:marschnervariance2}
var(\hat p_Y)=\left [ 1 - \rho^2 (1-\pi) \right ]
\frac{p_Y (1-p_Y)}{m},
\end{equation}
where $\rho$ is the correlation between $Y_i$ and $Z_i$, so as above
\[
\rho^2 = \frac{(p_{Y \mid Z}-p_Y)^2 p_Z}{p_Y (1-p_Y) (1-p_Z)}\quad \textrm{ and }\quad \pi = m/m_Z.
\]

\cite{vanlancker2020} extended this framework by allowing for different type of endpoints (e.g., continuous, binary and ordinal) and the incorporation of baseline measurements $\mathbf{X}_i$ that are correlated with $Y_i$ and whose effect can depend on treatment by building on the general framework of sequential augmented inverse probability weighting estimation (AIPW; see e.g., \cite{bang2005doubly}). 
As can be seen in Figure~\ref{fig_impact}, at the time of (early) trial analysis due to the COVID-19 outbreak, there will be three patient categories with observed data: (i) those who have completed the trial, (ii) those for whom baseline covariates and the intermediate outcome were observed at the time the trial was interrupted, and (iii) those who started but had their trial interrupted without any additional data observed besides baseline covariates.

For simplicity, we first consider the mean $\mu_r=E(Y_i|R=r)$ in a single treatment group $r$ ($r=0, 1$).
Although the outcome is only seen for the patients in cohort 1 (receiving the considered treatment $r$), under simple randomization and random recruitment the missing outcomes for the other patients can be unbiasedly predicted in large samples. In particular, an estimator for the average outcome $\mu_r$ can be obtained by:
\begin{enumerate}
	\item fitting a regression for the conditional mean of the outcome $Y_i$ given the baseline covariates $\mathbf{X}_{i}$ and the intermediate measurement $Z_i$ among the complete cases (i.e., patients in cohort 1) in the considered treatment group $r$ using a canonical generalized linear working model with intercept for the conditional mean of $Y_i$: $E(Y_i|R_i=r, \mathbf{X}_{i}, Z_i) = h_{r}(\mathbf{X}_{i}, Z_i, \boldsymbol{\eta}_0)$, where $h_{r}(\mathbf{X}_{i}, Z_i, \boldsymbol{\eta})$ is a known function, evaluated at a parameter $\boldsymbol{\eta}$ with unknown population value $\boldsymbol{\eta}_0$; for example, $h_{r}(\mathbf{X}_{i}, Z_i, \boldsymbol{\eta})=\eta_1+\eta_2Z_i+ \boldsymbol{\eta_3}^{\prime} \mathbf{X}_{i}$ for a continuous outcome $Y_i$ and $h_r(\mathbf{X}_{i}, Z_i, \boldsymbol{\eta})=\text{logit}^{-1}(\eta_1+\eta_2Z_i+\boldsymbol{\eta_3}^{\prime} \mathbf{X}_{i}$ for a binary outcome $Y_i$,
	\item using this regression model to predict the outcome for all patients in cohort 1 and 2 of the considered treatment group $j$ based on their observed baseline covariates $\mathbf{X}_{i}$ and their intermediate measurement $Z_i$ as $\hat{Y}_{ri}=h_r(\mathbf{X}_{i}, Z_i, \hat{\boldsymbol{\eta}})$,
	\item fitting a regression of these predictions $\hat{Y}_{ri}$ on the baseline covariates $\mathbf{X}_{i}$ among all patients in cohort 1 and 2 of the considered treatment group $r$ using a canonical generalized linear working model (with intercept) for the conditional mean of $\hat{Y}_{ri}$: $E(\hat{Y}_{ri}|R_i=r, \mathbf{X}_{i}) = g_r(\mathbf{X}_{i}, \boldsymbol{\zeta}_0)$, where $g_r(\mathbf{X}_i, \boldsymbol{\zeta})$ is a known function, evaluated at a parameter $\boldsymbol{\zeta}$ with unknown population value $\boldsymbol{\zeta}_0$; for example, $g_r(\mathbf{X}_{i}, \boldsymbol{\zeta})=\zeta_1+\boldsymbol{\zeta_2}^{\prime} \mathbf{X}_{i}$ for a continuous outcome $Y_i$ and $g_r(\mathbf{X}_{i}, \boldsymbol{\zeta})=\text{logit}^{-1}(\zeta_1+\boldsymbol{\zeta_2}^{\prime} \mathbf{X}_{i})$ for a binary outcome $Y_i$,
	\item using this regression model to predict $Y_i$ for all patients (also the other treatment group) based on their observed baseline covariates $\mathbf{X}_{i}$ as $\tilde{Y}_{ri}=g_r(\mathbf{X}_{i}, \hat{\boldsymbol{\zeta}})$.
\end{enumerate}

The estimator $\hat\mu_r$ for $\mu_r$ is obtained by taking the average of the predicted values $\tilde{Y}_{ri}$ over all patients.
Under random recruitment, this procedure will in general lead to an efficiency gain and higher power (compared to only using the primary outcomes; i.e., the dark blue cohort in Figure \ref{fig_impact}), without compromising the Type I error rate of the procedure in large samples, even when the adopted prediction models are misspecified \cite{bang2005doubly, rosenblum2015, vanlancker2020}. 

Define the indicator $C^Y_i$, which equals 1 if $Y_i$ is observed and 0 otherwise. Similarly, the indicator $C^Z_i$ equals 1 if $Z_i$ is observed and 0 otherwise.
Moreover, define $\hat{p}_R=\sum_{i=1}^{m_R} R_i/m_R$, $\hat{p}_{\bar{R}}=1-\hat{p}_R$, $\hat{p}_{C^Y|R}=\sum_{i=1}^{m_R} C^Y_iR_i/\sum_{i=1}^{m_R} R_i$ and $\hat{p}_{C^Y|\bar{R}}=\sum_{i=1}^{m_R} C^Y_i(1-R_i)/\sum_{i=1}^{m_R} (1-R_i)$. Then, the variance of the treatment effect estimator $\hat\mu_1-\hat\mu_0$ can be estimated as $1/m_R$ times the sample variance of the values 
$$\frac{R_iC^Y_i}{\hat{p}_R\hat{p}_{C^Y|R}}(Y_{i}-\hat Y_{1i})+\frac{R_iC^Z_i}{\hat{p}_R\hat{p}_{C^Z|R}}(\hat Y_{1i} - \tilde Y_{1i})+\tilde Y_{1i} - \hat\mu_1-\left(\frac{R_iC^Y_i}{\hat{p}_{\bar{R}}\hat{p}_{C^Y|\bar{R}}}(Y_i-\hat Y_{0i})+\frac{R_iC^Z_i}{\hat{p}_{\bar{R}}\hat{p}_{C^Z|\bar{R}}}(\hat Y_{0i} - \tilde Y_{0i})+\tilde Y_{0i} - \hat\mu_0\right),$$
for $i=1, \dots, m_R$ or via the nonparametric bootstrap. 

\section{External Auxiliary Information}
\label{sec:ext}

External information relevant to $\theta$ is the only way to overcome limitations of a given dataset (including internal information if available). In this section we present potential sources of such external information, as well as methods to incorporate this information in statistical inference.  

\subsection{Sources of External Auxiliary Information}
\label{sec:sources_ext}

Typical sources of additional external information include previously collected (historical) data, previous reports, and publications. Sometimes, this external information is available on the individual subject level. This situation may arise, e.g., in pharmaceutical development, based on another (historical or concurrent) study in the same development program, or placebo data from a different development program with the same indication from the same sponsor. Public data sources, e.g. from collaborative industry initiatives (i.e.  %
\hyperlink{transcelerate}{https://www.transceleratebiopharmainc.com}) 
or real-world observational data and electronic health records, may also qualify. More frequently, however, only summary-level data (such as descriptive statistics or published model results) are available for the external information. In principle, individual-level external data can provide richer information than external summary data, and different methods lend themselves to their inclusion.

In addition to external data sources, the Bayesian approach allows leveraging of expert opinion summarized into a prior distribution \cite{garthwaite2005statistical, o2006uncertain}. However, now we restrict our attention to the use of information coming from external data sources, and provide more details about the Bayesian approach in Section \ref{sec:BaysianMethods}.  

\subsection{Assessing and Dealing with Heterogeneity}

One important consideration for the inclusion of external information is how ``similar'' the external data source is to the trial that it is supposed to complement. This can range from an ideal situation, such as an identically designed ``sister trial'' concurrently conducted by the same sponsor in the same geographical regions, to much more difficult situations, such as an historical real-world data source to complement a modern clinical trial. Below we present  several important points, overlapping with meta-analytic guidelines (see \cite{cochrane}), to consider  (see also Table 1 in \cite{burger2021extcontr}).

\begin{itemize}
    \item {\bf Data age.} Time trends exist in clinical data \cite{nicholas2011trends}. They may stem from, for example, improvements in natural background care, but more often than not, the reasons for such trends cannot be discerned. If a common reference arm (often placebo) is available for both the external data and the trial that is to be augmented, it is important to check this for differences. 
    \item {\bf Population differences.} Differences in eligibility criteria, and/or in the actual population characteristics, should be explored, and adjusted for as appropriate.
    \item {\bf Endpoint definition.} It is not uncommon that clinical endpoints exist in different versions, sometimes even when referred to by the same name. E.g., composite endpoints may use different components; scales may use different anchor values; measurement devices may be differently calibrated; etc. Occasionally, if available, it is possible to use only those components of a composite score that are used in both the external data source and the trial to be augmented. On other occasions, one may attempt to ``translate'' one endpoint into the other if, e.g., a regression shows high correlation. 
    \item {\bf Treatment regimens.} If the external data source uses a different treatment regimen (dose and/or dosing frequency, administration route, etc.) than the trial at hand, additional steps may be required to include this data. For example, if dose linearity has been established, a different dose can be adjusted for in a straightforward way. In other cases, more sophisticated modeling techniques may be required, leveraging pharmacokinetic-pharmacodynamic (PKPD) or even physiologically-based pharmacokinetic (PBPK) principles.
    \item {\bf Geographical region.} The external data may not come from the exact same region of the world as the trial at hand. If there is heterogeneity in the treatment effect between regions, then a weighting scheme may be applied, or in extreme cases, only data from common geographical subregions may be used. The same may apply to different ethnicity groups recruited into the trial.
    \item {\bf Other aspects.} There are more aspects to be considered, e.g. permitted co-medications and other contextual conditions. This is especially important when combining real-world data with a clinical trial, as they may differ substantially.
\end{itemize}

The items above are potential sources of dissimilarity, which can be taken into consideration when known, e.g. by regression techniques. For example, if there is an association between participants' age and a study outcome, regression adjustment by age may reduce the impact of age differences between the main and an external study on a treatment effect under consideration. However, there may also be unknown reasons for dissimilarity. In the past decade, statistical concepts have been developed that measure to what degree two distributions are ``similar.'' If the data generating distribution of the external dataset is very similar to the one it should complement, we would like to include all the external information at face value. This situation is rare, but if it holds, a data pooling approach may be appropriate. For example, \cite{hua2021phase} describes a case where pooling is used for two concurrent ``sister trials'' from the same sponsor, to compensate for information loss due to the Covid-19 pandemic. If, on the other hand, the external dataset is completely dissimilar, it had better be discarded. In most cases, however, researchers encounter intermediate situations. A tuning mechanism should then allow to adjust the amount of information to be included, depending on the degree of similarity. More details on these measures and mechanisms are reported in Sections \ref{sec:AuxiliarySummaryStatistics} and \ref{sec:BaysianMethods}.

\subsection{Use of Auxiliary Summary Statistics}  
\label{sec:AuxiliarySummaryStatistics}
Some researchers have explored the use of auxiliary information available in the form of known population quantities, see \cite{Zenkova2017, Dmitriev2014}; others have used auxiliary information available in form of summary statistics such as sample means and standard errors, see \cite{Tarima2006, Tarima2007}. Asymptotic relative efficiency of the estimators proposed in \cite{Tarima2006} is derived in \cite{Albertus2022}.

Sometimes auxiliary information comes from an external data source with a somewhat different population, treatment, and/or endpoint definition, which can potentially add bias to the estimation procedure based on primary dataset. In \cite{Dmitriev2014} and \cite{tarima2020estimation}, the possibility of additional bias associated with the use of auxiliary information was mitigated by considering mean squared error minimization. In both papers the asymptotic distribution of estimators modified with auxiliary information is non-degenerate but different from normal. These asymptotic distributions produce shorter confidence intervals than the confidence intervals based on empirical data only. On the other hand, if the bias is strong, the effect of ``biased'' auxiliary information is suppressed and the estimator becomes asymptotically equivalent to the estimator without the use of additional information \cite{tarima2020estimation}.

The initial plan of enrolling $\omega_1,\ldots,\omega_n$ would lead to the MLE $\hat\theta_{(n)}$. Usually, this $\hat\theta_{(n)}$ is either unbiased $E(\hat\theta_{(n)})=\theta$, or asymptotically unbiased ($\lim_{n\to\infty}E(\hat\theta_{(n)})=\theta$) under certain regularity conditions of a central limit theorem (CLT). In the weight loss example, the primary outcome is two-year weight loss as assessed at physician visits ($Y_{i}$). When the pandemic struck, only $m$ patients had their two-year weight loss recorded. The primary outcome across these $m$ patients is aggregated into a summary statistic $$\hat\theta_{(m)} = \frac{\sum_{i=1}^m I( R_i=1) Y_{i}}{\sum_{i=1}^m I(R_i=1)} - \frac{\sum_{i=1}^m I(R_i=0) Y_{i}}{\sum_{i=1}^m I( R_i=0)},$$ which is an unbiased estimator of $\theta$. In addition to $\hat\theta_{(m)}$, its variance  $\widehat{var}\left(\hat\theta_{(m)}\right)$ is calculated on the pre-disruption data.
More generally, any other consistent estimator of $\theta$ can be used instead of $\hat\theta_{(m)}$ defined above.

Below we consider  external sources of auxiliary information available in the form of summary statistics. We also show how the method can be applied when individual data are available (since the summary statistics can be calculated).

\paragraph{External auxiliary information.} \label{external_summary} Suppose another study was conducted where a similar intervention was proposed, but the duration of the study was only $12$ months. This other study was published reporting averaged $12$ month weight loss as $$\hat\psi_{d(n_d)} = \frac{\sum_{i=1}^{n_d} I( R_{id}=1)Z_{id}}{\sum_{i=1}^{n_d} I(R_{id}=1)} - \frac{\sum_{i=1}^{n_d} I(R_{id}=0)Z_{id}}{\sum_{i=1}^{n_d} I(R_{id}=0)},$$ 
with $\widehat{var}\left(\hat\psi_{d(n_d)}\right)$. Obviously, the main sample $\omega_1^{(0)},\ldots,\omega_{n^{(0)}}^{(0)}$ can also be used to estimate the $12$ month effect, $$\hat\psi_{0(n_0)} = \frac{\sum_{i=1}^{n_0} I( R_{i0}=1)Z_{i0}}{\sum_{i=1}^{n_0} I(R_{i0}=1)} - \frac{\sum_{i=1}^{n_0} I(R_{i0}=0)Z_{i0}}{\sum_{i=1}^{n_0} I(R_{i0}=0)},$$ with $\widehat{var}\left(\hat\psi_{0(n_0)} \right)$.    
More generally, $\psi$ can be any other quantify estimable on external data. For example, $\psi$ can be defined not only as 12-month mean weight loss but also as 24-month mean weight loss, a rate of weight loss, 24-month median weight loss, etc. The methods presented in this section are applicable for many different external quantities. The important part is that this quantity should also be estimable with asymptotically normal estimators on both the external and main datasets.  Thus, $\psi$ is a population quantity estimated with $\hat\psi_{d(n_d)}$ on external data and estimated with $\hat\psi_{0(n_0)}$ on the main data.

\paragraph{Connection with Internal Auxiliary Information.} \label{internal_summary} 

The short-term endpoint $Z_{i}$ ($12$ month BMI reduction) is collected for some but not all patients: $m_Z-m$ who have a $12$ month assessment but do not have a $24$ month value for BMI reduction. These measurements are aggregated into a sample $12$ month summary measure $$\check\psi_{0(m_Z-m)} = \frac{\sum_{i=m+1}^{m_Z} I( R_{i0}=1)Z_{i0}}{\sum_{i=m+1}^{m_Z} I(R_{i0}=1)} - \frac{\sum_{i=m+1}^{m_Z} I( R_{i0}=0)Z_{i}^{(0)}}{\sum_{i=m+1}^{m_Z} I(R_{i0}=0)},$$ with  $var\left(\check\psi_{0(m_Z-m)}\right)$. 
The expectation of $\check\psi_{0(m_Z-m)}$, $E(\check\psi_{0(m_Z-m)}) = \psi +\delta$, is likely different from $\theta$. 
\paragraph{Method.} \label{Method_MMSE}
To avoid excessive subscripts, below we use ``hat'' and ``check'' respectively to denote estimates based on the main data source (pre-disruption data) and auxiliary estimates. Auxiliary data can be viewed as either internal or external auxiliary data  as described in Sections \ref{internal_summary} and \ref{external_summary}. 
Consider a class of estimators combining $\hat\psi_{0}$ (a pre-disruption estimate of $\psi$ on $\omega_1,\ldots,\omega_{m}$), $\check\psi$ (additional information based on a pre-disruption estimate of $\psi$ on $\omega_{m+1},\ldots,\omega_{m_Z}$, or external published study results), and $\hat\theta$ (pre-disruption primary outcome calculated on $\omega_1,\ldots,\omega_{m}$):
\begin{equation}\label{class_of_le}
\theta^{\Lambda} = \hat\theta + \Lambda\left(\hat\psi - \check\psi\right) = 
\hat\theta + \Lambda\hat \delta.
\end{equation}
Among the estimators $\theta^{\Lambda}$, the estimator 
\begin{equation}
\label{optest_MMSE_mult} 
\theta^{0}(\delta) = \hat\theta - cov\left(\hat\theta, \hat\delta\right)E^{-1}\left(\hat\delta\hat\delta^T\right) \hat\delta^T
\end{equation}
has the smallest mean squared error (MSE) 
\begin{equation}
\label{MSEoptest_MMSE_mult} 
MSE\left(\theta^{0}\right) = var\left(\hat\theta\right) - 
cov\left(\hat\theta, \hat\delta\right)E^{-1}\left(\hat\delta\hat\delta^T\right)cov\left(\hat\delta, \hat\theta\right),
\nonumber
\end{equation}
where $E\left(\hat\delta\hat\delta^T\right) = var\left(\hat\psi\right) + var\left(\check\psi\right) + \delta\delta^T$. With $\delta = 0$, $E(\theta^{\Lambda})=\theta$ $\forall \Lambda$, and
\begin{equation}
\label{optest_MVAR_mult} 
\theta^{0}(0) = \hat\theta - cov\left(\hat\theta, \hat\delta\right)cov^{-1}\left(\hat\delta\right) \hat\delta^T
\end{equation}
secures the smallest variance among $\theta^{\Lambda}$, see \cite{Tarima2006}, 
\begin{eqnarray}
var\left(\theta^{0}(0)\right) &=& var\left(\hat\theta\right) -
cov\left(\hat\theta, \hat\delta\right)var^{-1}\left(\hat\delta\right)cov\left(\hat\delta, \hat\theta\right).
\label{MSEoptest} 
\end{eqnarray}
The quadratic term in Equation (\ref{MSEoptest}) shows reduction in variance  when $\delta=0$. In practice, estimates of unknown variances and covariances are used to define a new estimator $\hat\theta^0(0)$. The estimator  $\hat\theta^0(0)$ is asymptotically equivalent to $\theta^{0}(0)$.

An interesting characteristic of $\theta^{0}(\delta)$ is that if $\delta$ is different from zero, the impact of auxiliary information ($\check\psi$) is suppressed at large samples, and $\theta^{0}(\delta)$ and $ \hat\theta$ become asymptotically equivalent. This protects the inference against an auxiliary information and main data conflict. 

The values of $\delta$ and, consequently, $\theta^{0}(\delta)$ are unknown in practice. When consistent estimates of unknown variances and covariances are used the large sample properties do not change. The value of $\delta$ is also unknown and also should be estimated. When $\hat\delta$ is applied, a new estimator, $\theta^{0}(\hat\delta)$, is generated. When $\delta \ne 0$,  $\theta^{0}(\hat\delta)$ and $\hat\theta$ are asymptotically equivalent.
If $\delta = 0$, then $\sqrt{n}\left(\theta^{0}(\hat\delta)-\theta\right)$ is no longer normally distributed, but still converges to a stationary random variable, which allows the calculation of narrower confidence intervals than the confidence intervals based on the normal asymptotics of $\hat\theta$. The distribution of $\sqrt{n}\left(\theta^{0}(\hat\delta)-\theta\right)$ can be estimated using parametric bootstrap resampling. 
Then a confidence interval can be constructed. Further, if in this parametric bootstrap resampling, we impose $\theta=0$, then asymptotic bootstrap p-values can be calculated.

\subsection{Use of Bayesian Methods}
\label{sec:BaysianMethods}
Bayesian approaches can be considered to be an additional tool to mitigate clinical trial disruptions and the consequent loss of information. Borrowing relevant external information can be achieved by defining informative prior probability distributions. Such prior distributions convey a summary of the available knowledge concerning the parameter(s) of interest, and can then be updated with the evidence collected during the current trial following the basic rules of probability.
However, as mentioned above, a concern when incorporating external information is the potential for heterogeneity, which can arise if the parameters of the external and the current trial data likelihood differ. Heterogeneity can severely affect frequentist operating characteristics, particularly when the external trial sample size is large compared to that of the current trial. Section \ref{sec:sources_ext} has reviewed potential known sources of heterogeneity, and suggested approaches to deal with such discrepancies. Often, such an assessment is further corroborated by sensitivity analyses and/or the adoption of prior specifications which allow the discounting of external information based on a data-driven assessment of the degree of similarity.

\paragraph{Commensurability.} 
To quantify the degree of similarity between external information and available data, commensurability parameters have been defined in the literature. Examples can be found in a bridging study context by \cite{takeda2015}, which used a class of commensurate prior distributions (\cite{hobbs2011,hobbs2012}), or by adopting a meta-analytic approach~\cite{schmidli2014}. \cite{ollier2020} proposed a parameter, using the Hellinger distance between the two normalized likelihoods:
\small
\begin{equation} 
\Delta^2 \left(D_{d(n_d)},D_{0(n_0)})\right)= 
\frac{1}{2} \displaystyle\int \left(\sqrt{\frac{L(\boldsymbol\theta ; D_{0(n_0)})^{\min\left(1,\frac{n_d}{n_0}\right)}}{\int L(\boldsymbol\theta ; D_{0(n_0)})^{\min\left(1,\frac{n_d}{n_0}\right)}d\boldsymbol\theta}}-\sqrt{\frac{L(\boldsymbol\theta ; D_{d(n_d)})^{\min\left(1,\frac{n_0}{n_d}\right)}}{\int L(\boldsymbol\theta ; D_{d(n_d)})^{\min\left(1,\frac{n_0}{n_d}\right)} d\boldsymbol\theta}}\right)^2 d\boldsymbol\theta,
\label{distance_Ollier}
\end{equation}
where $n_0$ and $n_d$ refer to the sample size of the actual trial and of the $d$th external trial, respectively. The likelihood containing more data is raised to a factor $\leq 1$ to allow it to be comparable to the one that contains
less information. The commensurability parameter can be then defined as $\Delta^c$, with $c \in R^+$. The advantage of this definition is that $\Delta$ is bounded between 0 and 1, providing an easy interpretation of the degree of similarity ($1-\Delta$). 
A few extensions of Equation ~\ref{distance_Ollier} have been proposed, for example by adding a weakly informative prior to both likelihoods to stabilize the computation~\cite{ollier2021}. Finally, a marginal version of  Equation ~\ref{distance_Ollier} can be used to compare the distribution of each model parameter separately.

\paragraph{Bayesian approaches for incorporating external information.}
The commensurability parameter defined in Equation \ref{distance_Ollier} \cite{ollier2020} can be used to control the amount of external information included in the analysis of the current study. Indeed, as previously mentioned, a commensurability parameter is at the core of various prior specifications which aim to gradually discount potentially heterogeneous external information.
We focus in the following on the power prior (see e.g. \cite{ibrahim2000,ibrahim2015}), and the meta-analytic combined or predictive prior \cite{neuenschwander2010,schmidli2014} approaches, assuming availability of one external data-source. Generalizations to multiple external data-sources are also possible and a more comprehensive review can be found in, e.g, \cite{neuenschwander2020}.

The power prior (see, e.g., \cite{ibrahim2000,ibrahim2015}) is obtained by raising the external data likelihood to a power $\lambda \in [0,1]$. $\lambda$ can be interpreted as the fraction of external samples retained when computing the posterior, and thus, the prior for the current study. Formally, the power prior is given by $g_{\theta|\mathbf{Y}_d}(\theta|\mathbf{Y}_d) \propto g_{\theta}(\theta) L(\theta; \mathbf{Y}_d )^\lambda$. The power parameter $\lambda$ may be fixed \textit{a priori}, estimated by assigning a prior to $\lambda$, via an empirical Bayes approach~\cite{gravestock2017}, or using the commenurability parameter defined in Equation \ref{distance_Ollier} \cite{ollier2020}. If estimated, it is recommended to divide the power prior by its normalizing constant (conditional on $\lambda$), i.e., $C(\lambda)=\int g_{\theta}(\theta) L(\theta; \mathbf{Y}_d )^\lambda d\theta$. This satisfies the likelihood principle and typically leads to improved behavior in terms of borrowing \cite{duan2006,neuenschwander2009}. Extensions of the power prior to regression models require raising the whole external data likelihood to the parameter $\lambda$: Formally, assuming the availability of external data $D_{d(n_d)}$, comprising, e.g., a matrix of covariates $\mathcal{X}_{d}$ and a vector of treatment assignmemts $\mathbf{R}_{d}$, the power prior is given by $g_{\boldsymbol\theta|D_{d(n_d)}}(\boldsymbol\theta |D_{d(n_d)}) \propto g_{\text{\boldmath$\theta$}}(\text{\boldmath$\theta$}) L(\text{\boldmath$\theta$};D_{d(n_d)} )^\lambda$. Note that the power prior is not tailored to deal with situations where borrowing is desired only for a subset of parameters. A potential solution consists in approximating the marginal posteriors for the parameters of interest from the external data analysis with suitable densities and inflating their variances by an appropriately chosen $\lambda$ (which may vary for each parameter). The latter approach is also suitable if only summary statistics are available from the external data source and will be exemplified in the case-study analysis.

Alternative approaches to borrow information from external data while accounting for potential heterogeneity between the current and external data sources, include the meta-analytic combined (MAC) and meta-analytic predictive (MAP) approaches  \cite{neuenschwander2016}. The meta-analytic combined approach takes advantage of a hierarchical formulation of the model i.e., it assumes that the parameter of the current and the external trial are exchangeable, but not equal, and drawn in turn from a common distribution. This distribution is typically a normal distribution with mean parameter $\xi$ and variance $\tau^2$, where $\tau^2$ quantifies the variability to be expected between studies. The parameter $\tau^2$ can be fixed \textit{a priori}, or, more consistently with the Bayesian approach, a prior distribution can be assigned to it. 
Multi-parameter models, e.g., models including patient-level covariates, can be handled in a  straightforward manner and only require the specification of the common prior linking the external and current study parameters. A potential advantage of the MAC over the power prior approach is that different degrees of commensurability can be assumed for each parameter without further model adaptations. For a given $\tau^2$ and a uniform improper prior for $\xi$, inferences equivalent to those of the MAC approach can be obtained with the use of the meta-analytic predictive (MAP) prior. Here, the prior is obtained by deriving the predictive distribution for the parameter of the current study based on the above-mentioned hierarchical model. 
The MAP approach is typically preferred in situations where the prospective use of historical is desired, i.e. when external information is available at the time of designing the current trial \cite{neuenschwander2016}. Depending on the sample size of the external study, and the value or prior distribution assigned to $\tau^2$, heterogeneity can still be of concern. The robust MAP prior mitigates this issue by defining a new prior based on the mixture of the MAP prior and a robust more dispersed distribution \cite{schmidli2014,kass1995}. 

The parameter $\delta$ in the power prior approach and the parameter $\tau^2$ in the meta-analytic approach 
can be functionally related in the case of a single external study and under a monoparametric `normal prior - normal outcome model'. Indeed, it has been shown that in such a situation $\lambda=(1+ 2 n_d \tau^2/\sigma_d^2)^{-1}$
under a uniform improper prior for $\xi$ \cite{chen2006}, where $\sigma_d^2$ is the variance of the external data outcome. An equivalent relationship between the power prior parameter and the heterogeneity parameter in the meta-analytic approach has been shown for regression models \cite{chen2006}. Such a relationship offers additional insight on how the proportion of external samples $\lambda$ incorporated into the analysis can be seen as a function of both the between trial heterogeneity $\tau^2$ and the external sample size $n_d$ \cite{neuenschwander2020}. 

\paragraph{Impact of incorporation of external information on frequentist operating characteristics.}
The impact of borrowing of external information on frequentist operating characteristics should be typically quantified via simulation, particularly if external information is discounted based on the observed degree of heterogeneity.  
 The work of  \cite{viele2014} explores the consequences of incorporating external information on type I error rates, power, and mean squared error, for several borrowing approaches. The results highlight the benefits that can be achieved if external information is consistent with that collected from the current trial (reduced MSE, increased power and reduced type I error), but losses can be incurred otherwise. Their investigation  concentrates on borrowing of information for the control arm, but the conclusions can be generalised. \cite{kopp2020} provides a formal proof of the impossibility to achieve power gains without inflation of maximum type I error rate if a uniformly most powerful (unbiased) test exists, even when external information is adaptively discounted based on the observed degree of prior-data conflict. In other words, inflation of type I error rates will be observed unless prior assumptions concerning the degree of commensurability between current and external information can be taken into account when specifying the data-generating process on which computation of the trial's operating characteristics is based. Data-dependent estimation of commensurability will only mitigate the issue, typically avoiding an extreme inflation.
It has to be noted, however, that appropriately selected external data should, in principle, contain valuable information about the current trial. 
Trust in the selected external information may thus motivate a `reasonable' - and appropriately quantified analytically or by simulation - inflation of type I error rates, as stated in the FDA guidance on the use of Bayesian Statistics for Medical Device Clinical trials \cite{fda}.

\section{Illustrative Application}
\label{sec:application}

This section applies the aforementioned approaches on the use of auxiliary information for improving the planned ANCOVA analyses for pBMI and zBMI outcomes in the dataset introduced in Section 2.2 above. 

\subsection{Models}

The adopted analysis model has the following form
\begin{equation}
zBMI3_{id}   =   \theta_{0d} + \theta_{1d}  zBMI1_{id} + \theta_{2d} X_{id} + \theta_{3d} R_{id} + \epsilon_{id},  \
\epsilon_{id} \sim  N(0, \sigma_{d}^2),
\label{eq:ancova}
\end{equation}
where $d=\{0,...,D\}$ is the trial index, $i=1,...n_d$ is the observation index, $\theta_{1d}, \theta_{2d}$, and $\theta_{3d}$ are the effect of the baseline measurement, the additional covariate, and the treatment effect, respectively, on the primary outcome in trial $d$. The same model is also fit by substituting the percentile variables $pBMI3_{id}$ and $pBMI1_{id}$  to the z-scores  $zBMI3_{id}$ and $zBMI1_{id}$, respectively, under the assumption that common regularity conditions hold and the MLE of regression coefficients is approximately normal.  

The results are shown in Tables~\ref{tab:results1} (frequentist approaches) and \ref{tab:results2} (Bayesian approaches), as well as Figure \ref{fig:resplots} and Supplementary Figure \ref{fig:resplots_p} (all approaches). We first illustrate the application of the two approaches using \textbf{internal auxiliary information} as described in Section \ref{sec:int}. 
The first approach, denoted by \textbf{DReg} in Table~\ref{tab:results1}, implements the double regression method of \cite{engel1991} and \cite{galbraith2003} with the first regression model (\ref{eq:doubleregression1}) relating the intermediate measurement BMI2 to treatment and the baseline measurement BMI1 for all patients with BMI2 data, and the second regression (\ref{eq:doubleregression2}) relating the outcome BMI3 to treatment, BMI1 and BMI2 for all patients with BMI3 data. The second approach, denoted by \textbf{AIPW} in Table~\ref{tab:results1}, implements the algorithm suggested by Van Lancker et al. that is robust against model misspecification. Here, in each treatment arm separately, a model for BMI3 is fitted on the covariate $X$, the baseline measurement BMI1 and the intermediate measurement BMI2 in Step 1 of the algorithm. The model in Step 3 of the algorithm includes the covariate $X$ and the baseline measurement BMI1. 

Incorporation of \textbf{external auxiliary information} is exemplified under the assumption that external trial data are available either in the form of summary statistics for the primary endpoint, i.e., BMI percentile/z-score reduction at 24 months in the treatment versus the control group, or as individual patient data (IPD).  To account for potential baseline and covariate effect differences between the current and external trials, interest is placed on borrowing information on the treatment effect parameter only. Both frequentist and Bayesian approaches are presented.

Two frequentist methods, denoted by \textbf{MVAR and MMSE}, for incorporating external auxiliary information are presented.  Statistical details on MVAR and MMSE methods for hypothesis testing and confidence interval estimation were presented in Section \ref{sec:AuxiliarySummaryStatistics}.

Three Bayesian approaches are illustrated. The first Bayesian approach is a hierarchical model which jointly estimates all model parameters from individual patient data and assumes
\begin{eqnarray}
\theta_{3d} \sim N(\xi, \tau^2), \  \xi \sim N(0,10^2), \ \tau \sim HN(0, \sigma_\tau^2).
\label{eq:priors}
\end{eqnarray}
This approach is therefore denoted \textbf{Hierarchical}. The value of $\sigma_\tau$ controls the amount of borrowing and, following the guidelines in \cite{neuenschwander2020}, a mildly informative assumption concerning its value is adopted: $\sigma_{unit}/4$, where $\sigma_{unit}$ represents the unit-information standard deviation for the treatment effect parameter (see, e.g., \cite{rover2021}), as estimated in the external dataset.
In the \textbf{Power} approach, the marginal posterior distribution for the treatment effect parameter of the external trial is first computed, and then used as a prior for the analysis of the current trial. Robustness is achieved by inflation of the prior variance according to the commensurability parameter $\Delta$ based on \ref{distance_Ollier}. Assuming approximate normality of this marginal posterior distribution, the resulting power prior is of the form $N[E(\theta_{3d}| D_{d(n_d)}), var(\theta_{3d}| D_{d(n_d)})/(1-\Delta)^2]$, where $var(\theta_{3d}| D_{d(n_d)})$ is computed multiplying the unit information by the number of missing data. Note that the Power approach can also be applied to summary statistics from the external data-source. In both the hierarchical and the Power approach independent $N(0,10^2)$ priors are assigned to $\theta_{0d}$, $\theta_{1d}$ and $\theta_{2d}$, while a Half-$t_3(0,10^2)$ prior distribution is assigned to the residual standard deviation $\sigma_{d}$.
The third Bayesian approach, denoted \textbf{MAC}, is analogous to that of meta-analytic studies, i.e., frequentist estimates of internal and external trial level parameters are independently obtained at a first stage, while the common prior and shrinkage trial-level parameter estimates are obtained at a subsequent second stage according to the MAC model described in Section \ref{sec:BaysianMethods}. Priors for the second stage parameters are analogous to the hierarchical approach, see Equation \ref{eq:priors}.

\subsection{Results}

Analysis outputs from the application of the frequentist approaches described above are shown in Table \ref{tab:results1}.  For comparison, results from a complete case analysis based only on data from patients with BMI3 data are also shown. Estimates, standard errors, and two-sided $p$-values are displayed for testing the null hypothesis that the treatment effect parameter at 24 months is equal to zero. 

Both approaches using internal data (DReg  and AIPW) lead to overall very similar results. Both methods demonstrate efficiency gain relative to the complete cases analysis, with slightly greater gain for the AIPW method. This efficiency gain increases with increase of the correlation between zBMI2 and zBMI3.

The MVAR method for incorporating external data is equivalent to simple pooling of two data sources together to build an estimator with the smallest variance. This is why this method is the best (unbiased and the smallest standard error) when there is no conflict between the main and external datasets. But a strong conflict between them leads to a biased estimate of the treatment effect: see MVAR, Ext-Full (SC) simulation setting in Table \ref{tab:results1}. To suppress the undesired impact of a conflict between the datasets, the MMSE procedure minimizes the mean squared error and the conflict between the main and external datasets is ``adaptively'' detected and suppressed: see MMSE, Ext-Full (SC) simulation settings in Table \ref{tab:results1}. 

\begin{figure}
    \includegraphics[width=1\textwidth]{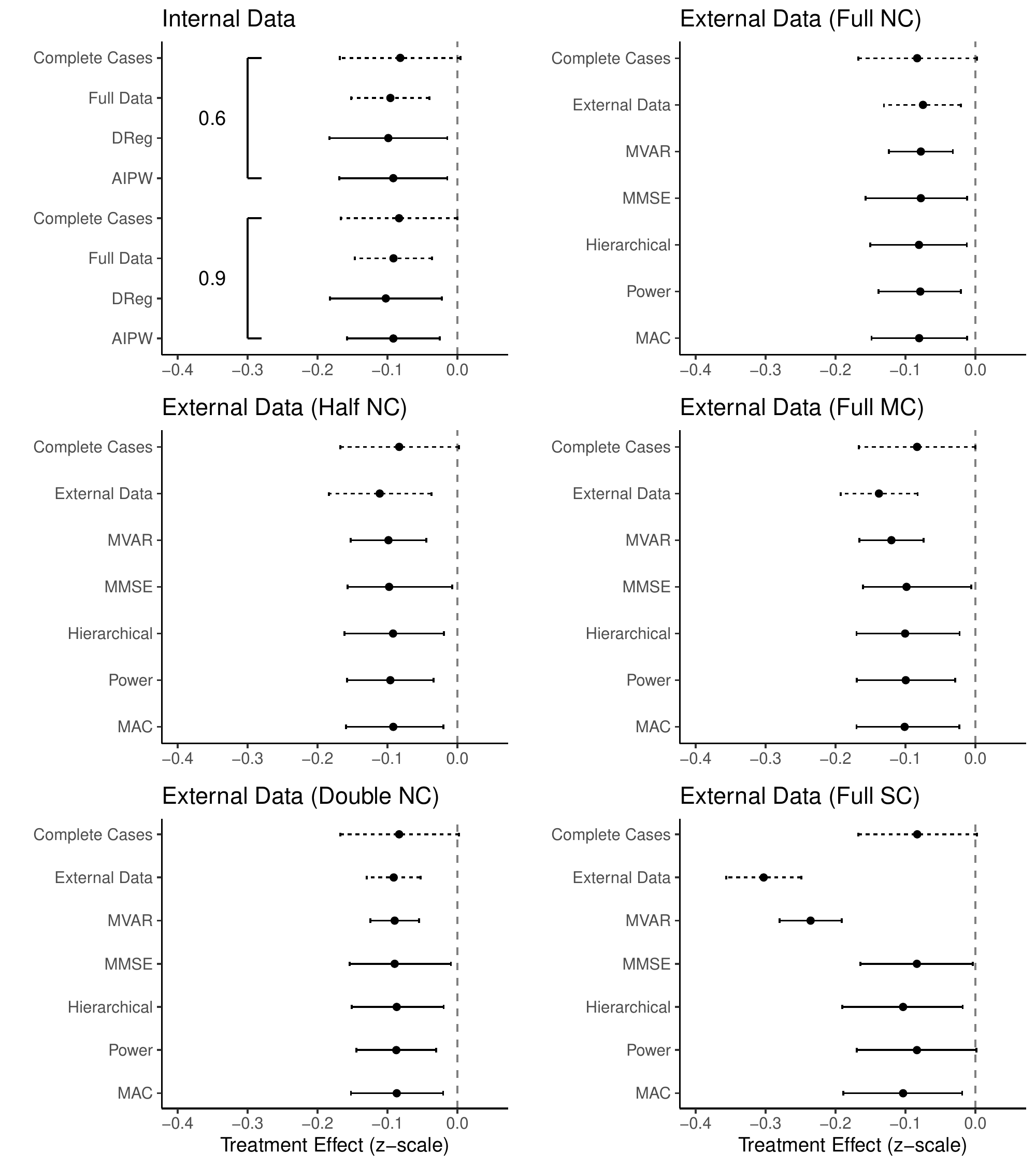}
    \caption{Analysis of the zBMI values: Estimates and 95\% intervals for the treatment effect parameter at 24 months. Confidence intervals are shown for frequentist methods, while symmetric credible intervals are shown for Bayesian methods. In the case of methods leveraging external information, results with incorporation of external data as well using only the complete cases without any borrowing are reported. For the internal methods, the result based on the full data-set prior to setting any value to missing (Full data) is included for illustrative purposes. The correlation between zBMI2 and zBMI3 is equal to 0.9 unless otherwise specified.}
    \label{fig:resplots}
\end{figure}

Thus, MMSE is robust to the conflict, but this robustness has its price. If there is no conflict between the datasets, the MMSE based method leads to a wider confidence interval than the MMSE procedure, (Figure \ref{fig:resplots}). Interestingly, the asymptotic distribution of the MMSE-based treatment effect estimator is stationary but not normal under $H_0$. Consequently, the P-value is calculated using this non-normal distribution.

\begin{table}[ht!]
\centering
 \begin{tabular}{|ll|cc|cc|} \hline
\multirow{2}{*}{Method}  & Auxiliary & \multicolumn{2}{c|}{pBMI} & \multicolumn{2}{c|}{zBMI} \\
        & information & Effect (SE) & $p$-value & 
        Effect (SE) & $p$-value\\ 
        \hline
 \multicolumn{6}{l}{}\\
 \multicolumn{6}{l}{\textbf{Use of Internal Information: Correlation between zBMI2 and zBMI3 = 0.6}}\\ \hline 
Complete cases &  None &
 -0.0124 (0.0067) & 0.0637 
 &  -0.0817 (0.0440) & 0.0647 
\\  \hline
DReg & Int-$\mathbf{BMI2}_{1:m_Z}$, & 
    -0.0139 (0.0061) & 0.0222 
    &
	-0.0988 (0.0430) & 0.0217 
	\\   
& $\{\mathbf{X},\mathbf{BMI1}\}_{(m+1):m_Z}$ & 
     &    & & \\   
AIPW
& Int-$\mathbf{BMI2}_{1:m_Z}$, & -0.0148 (0.0059) &  0.0122 
&  -0.0917 (0.0394) & 0.0199  
\\  
& $\{\mathbf{X}, \mathbf{BMI1}\}_{1:m_R}$ & 
&  &  & \\  
 \hline
   \multicolumn{6}{l}{}\\
 \multicolumn{6}{l}{\textbf{Use of Internal Information: Correlation between zBMI2 and zBMI3 = 0.9}} \\ \hline 
Complete cases & None &  -0.0110 (0.0060) & 0.0706 & 
 -0.0834 (0.0425) & 0.0515 
 \\ \hline 
DReg &  Int-$\mathbf{BMI2}_{1:m_Z}$,&
    -0.0110 (0.0048) & 0.0211 & 
    -0.1024 (0.0409) & 0.0122 
    \\
&  $\{\mathbf{X}, \mathbf{BMI1}\}_{m+1:m_Z}$ &
    &  &  & \\    
AIPW
& Int-$\mathbf{BMI2}_{1:m_Z}$,& -0.0111 (0.0047)  &  0.0176 &  
-0.0916 (0.0339) & 0.0069  
\\  
& $\{\mathbf{X}, \mathbf{BMI1}\}_{1:m_R}$&  &  & 
 & \\  
\hline

    \multicolumn{6}{l}{}\\
 \multicolumn{6}{l}{\textbf{Use of External Information}} \\ \hline 
Complete cases & None &  -0.0110 (0.0060) & 0.0706 & 
 -0.0834 (0.0425) & 0.0515 
 \\ \hline 

&   Ext-Full (NC)  & -0.0094(0.0035) & 0.0067 & 
-0.078(0.0233) & 0.0008 
\\ 
  &  Ext-Double (NC) & -0.0111(0.0027) & <.0001 & 
  -0.0897(0.0177) & <.0001 
  \\ 
  MVAR &   Ext-Half (NC) & -0.0133(0.0039) & 0.0005 & 
  -0.0986(0.0275) & 0.0003 
  \\ 
  &   Ext-Full (SC) & -0.0277(0.0035) & <.0001 & 
  -0.2356(0.0227) & <.0001 
  \\ 
  &   Ext-Full (MC) & -0.0141(0.0034) & <.0001 & 
  -0.1201(0.0234) & <.0001 
  \\ \hline
   &   Ext-Full (NC)  & -0.0094(0.0037) & 0.0171 & 
   -0.078(0.0355) & 0.0419 
   \\ 
  &   Ext-Double (NC) & -0.0111(0.0029) & 0.0014 & 
  -0.0897(0.0337) & 0.0235 
  \\ 
  MMSE &   Ext-Half (NC) & -0.0133(0.0042) & 0.0033 & 
  -0.0976(0.0378) & 0.0121 
  \\ 
  &   Ext-Full (SC) & -0.0136(0.0067) & 0.0667 & 
  -0.0837(0.0413) & 0.0460 
  \\ 
  &   Ext-Full (MC) & -0.0141(0.0038) & 0.0010 & 
  -0.0985(0.0404) & 0.0119 
  \\    \hline  
 \end{tabular}
 \caption{\label{tab:results1} Frequentist approaches: Estimates (standard errors) and $p$-values for the treatment effect parameter at 24 months. The complete cases analysis is based on frequentist inference for the ANCOVA model in Equation \ref{eq:ancova}. Results for different frequentist approaches leveraging either internal (Int) or external (Ext) trial data are reported. For the external data-sets, different levels of observed heterogeneity are assumed:  No conflict (NC), Moderate conflict (MC), and Strong conflict (SC). In addition, different sample sizes for the external trial data are considered: Same size as the main study but without missing data (Full), double size with respect to Full (Double), or half size with respect to Full (Half).}
\end{table}

\begin{table}[ht!]
\setlength{\tabcolsep}{4pt}
\centering
 \begin{tabular}{|ll|cc|cc|} 
  \multicolumn{6}{l}{\textbf{Use of External Information}}\\ 
 \hline
 \multirow{2}{*}{Method}  & Auxiliary & \multicolumn{2}{c|}{pBMI} & \multicolumn{2}{c|}{zBMI} \\
        & information  & Post.Mean (SD) & pr($\theta \geq 0$|D) & 
        Post.Mean (SD) & pr($\theta \geq 0$|D) 
        \\
        \hline \hline
Complete cases & None  & -0.011 (0.0062) & 0.0371 
 & -0.0833 (0.0429) & 0.0261 
 \\ \hline
\multirow{5}{*}{Hierarchical} & Ext-Full (NC) & -0.0095 (0.004) & 0.0094 
& -0.0807 (0.0346) & 0.0118 
\\ 
   & Ext-Double (NC) & -0.0111 (0.0033) & 0.0012 
   & -0.0868 (0.0325) & 0.0077 
   \\ 
   & Ext-Half (NC) & -0.0134 (0.0042) & 0.0005 
   & -0.092 (0.0356) & 0.0084 
   \\ 
   & Ext-Full (SC) & -0.0234 (0.0058) & 0.0001 
   & -0.1034 (0.0439) & 0.0091 
   \\ 
   & Ext-Full (MC) & -0.014 (0.0038) & 0.0006 
   & -0.1005 (0.0373) & 0.0064 
   \\  \hline
  \multirow{5}{*}{Power} & Ext-Full (NC) & -0.0098 (0.0046) & 0.0158 
  & -0.0788 (0.0299) & 0.0042 
  \\ 
   & Ext-Double (NC) & -0.011 (0.0043) & 0.0050 
   & -0.0874 (0.029) & 0.0013 
   \\ 
   & Ext-Half (NC) & -0.0129 (0.0046) & 0.0025 
   & -0.0959 (0.0313) & 0.0011 
   \\ 
  & Ext-Full (SC) & -0.011 (0.006) & 0.0329 
  & -0.0836 (0.0435) & 0.0273 
  \\ 
  & Ext-Full (MC) & -0.0128 (0.0048) & 0.0039 
  & -0.0996 (0.0362) & 0.0030 
  \\  \hline
  \multirow{5}{*}{MAC} & Ext-Full (NC) & -0.0095 (0.004) & 0.0082 
  & -0.0804 (0.0345) & 0.0109 
  \\ 
   & Ext-Double (NC) & -0.0111 (0.0034) & 0.0013 
   & -0.0867 (0.0327) & 0.0073 
   \\ 
   & Ext-Half (NC) & -0.0134 (0.0041) & 0.0005 
   & -0.0917 (0.0351) & 0.0066 
   \\ 
   & Ext-Full (SC) & -0.0231 (0.0057) & 0.0001 
   & -0.1034 (0.0435) & 0.0086 
   \\ 
   & Ext-Full (MC) & -0.014 (0.0038) & 0.0004 
   & -0.1012 (0.0371) & 0.0064 
   \\ 
 \hline
 \end{tabular}
 \caption{\label{tab:results2} Bayesian approaches: Posterior means (posterior standard deviations) and posterior probabilities of the treatment effect parameter at 24 months being greater or equal to zero. The complete cases analysis is based on Bayesian inference for the ANCOVA model (see Equation \ref{eq:ancova}). Results for different Bayesian approaches leveraging external (Ext) trial data are reported. Different levels of observed heterogeneity are assumed:  No conflict (NC), Moderate conflict (MC), and Strong conflict (SC). In addition, different sample sizes for the external trial data are considered: Same size as the current study but without missing data (Full), double size with respect to Full (Double), or half size with respect to Full (Half). $D$ denotes the complete cases of the current trial $D_{0(m)}$ for the complete cases analysis, or both $D_{0(m)}$ and $D_{d(n_d)}$ in all other analyses which incorporate external information.}
\end{table}

Results of the three Bayesian approaches outlined above are reported in Table \ref{tab:results2}. Here, posterior means, posterior standard deviations, and posterior probabilities of the treatment effect being greater or equal to zero, are shown. For consistency with the frequentist methods for incorporating external information, results are reported for the same simulation settings (see section \ref{sec:casestudy}). Posterior standard deviations are always reduced by the borrowing approaches as compared to the complete cases analysis, with the sole exception of the strongly conflicting scenario on the zBMI scale. In this case, uncertainty is slightly increased. In the power approach, external information is effectively discarded ($\Delta^2=0.98$). The small increase in variability is likely attributed to the numerical variability of the MCMC samples. In the MAC and Hierarchical approaches, this small increase in uncertainty is accompanied by a non-negligible increase in negative bias leading to decreased posterior probability that the parameter is greater or equal to zero. 
The Hierarchical and MAP approaches lead to similar results and confirm appropriateness of approximate normality with known variance assumptions for estimating the treatment effect. 

 Point and interval estimates (confidence or credible intervals as appropriate) on z-scale for all approaches can be visually compared in Figure \ref{fig:resplots}. Supplementary Figure \ref{fig:resplots_p} also displays point and interval estimates but on p-scale. The figures visualize how different external methods combine the complete cases and external data estimates into a single random quantity summarized by a point and an interval estimates.

\section{Discussion}
\label{sec:Discussion}

The aim of this paper was to collate and discuss methods for coping with information loss due to unplanned clinical trial disruptions among which the recently caused by the COVID-19 pandemic. We focused on methods that make use of auxiliary sources of data, either from the interrupted trial itself or from external sources. Broadly speaking, the main mechanism for the use of trial-internal data is to retrieve information about the primary endpoint from correlated measures, such as earlier realizations of the same endpoint, other correlated endpoints, or – under certain conditions – even baseline data. Baseline data, if related to the outcome, can improve precision and power of statistical procedures \cite{benkeser2021improving}. Justifying the inclusion of external information typically phrased in terms of similarity (or conversely, heterogeneity) between the disrupted trial and the external source. Statistical parameters such as commensurability or between-trial variability can quantify the degree of similarity and control the amount of borrowing (or conversely, discounting) of the external information.

A full Bayesian approach would assign a distribution to a commensurability parameter, as exemplified for the hierarchical and MAC approaches. We note, however, that if the number of studies is small, results will tend to be quite sensitive to prior assumptions concerning $\tau^2$, as the data carry only limited information concerning its value. The case-study example has shown strong similarity between the hierarchical and MAC approaches in terms of inferences for the current study treatment effect. The MAC approach can be advantageous in terms of computational stability and is appropriate when  the variance is known and approximate normality of the first stage estimators holds (\cite{lunn2013}). These are, in practice, common and appropriate assumptions for typical parameters if trial sample sizes are not very small, as in the current case-study example. The hierarchical and MAC approaches are typically employed in the presence of concurrent external data, i.e., data collected at the same time as the current trial \cite{neuenschwander2016}. In contrast, the Power approach illustrated in the case-study is more tailored to situations where external data is available at the time of designing the current trial, and makes use of a commensurability measure to quantify and control the proportion of external samples to be included in the analysis of the current data.

It is worth mentioning a few additional methods not covered in the space of this paper. Methods that can help in situations where the external data source is not readily comparable to the disrupted trial. For example, if the external source has different inclusion criteria, then a subset could be taken that most resembles the trial at hand. Matching on baseline characteristics could enhance comparability. The propensity to be included in a trial and/or to receive a certain treatment could be computed for non-randomized external data, and matching could be performed on the propensity scores (see e.g., \cite{rosenbaum2010design, austin2006comparison}). Information could be weighted based on inverse probability of treatment weighting (see e.g., \cite{rosenbaum2010design, austin2006comparison}). Other approaches may help address potential concerns about privacy and consent, given that external data sources are often originally not intended to be used for supplementing another (disrupted) trial. One recently popularized approach in this context is to generate synthetic data, resembling the real (external) data in relevant ways without compromising patient privacy or consent (\cite{synthdata1}, \cite{synthdata2}). Importantly, for all these methods as well as the ones described in this paper, a thorough understanding of their benefits, limitations, and potential pitfalls is required before proceeding to implementation. This necessitates thorough discussions and often scenario simulations, potentially in interaction with regulatory bodies.

The frequentist MVAR procedure is useful when the main and external data are drawn from the same population. It is possible that the external dataset does not have the same primary study outcome recorded. For example, 24-month change ($\theta$) in BMI may not be available from the external data, but another quantity is available: for example, 12-month change in BMI ($\psi$) is estimated in both datasets: $\hat\psi$ in the main dataset and $\check\psi$ in the external dataset. It is necessary that this other quantity is correlated with the main study outcome. Furthermore, it should be estimable without bias on both the main and external datasets: $E(\hat\psi) - E(\hat\psi) = 0$. This is typically a direct consequence of the fact that both datasets are drawn from the same population. In practice, however, there is no guarantee that both datasets are drawn from the same population. If this happens, $E(\hat\psi) - E(\hat\psi) = \delta \ne 0$.

This is where the MMSE helps: it protects estimation based on both main and external data from the situation when populations differ. Different populations for the main and externals datasets lead to bias ($\delta$) when the same quantities are estimated on these datasets. This discrepancy, if it exists, is detected by the MMSE procedure (whereas MVAR does not detect it) and its impact on estimation and hypothesis testing is adaptively suppressed. The MMSE procedure suppresses external information if the magnitude of the bias converges to zero or another constant at a rate slower than $1/\sqrt{\min(n_0,n_1)}$ or diverges.

If bias is absent ($\delta = 0$), for example, the main and external populations are the same, MVAR is preferred to MMSE. MVAR has higher power in hypothesis testing and leads to more accurate estimation than the MMSE. The MMSE procedure still outperforms estimation and hypothesis testing based on main data only. The combined estimator based on the MMSE procedure is no longer asymptotically normal, but its distribution continues to converge to a stationary distribution which can be used for hypothesis testing purposes. A parametric bootstrap can be used for building confidence intervals for the MMSE estimators. These confidence intervals are shorter than the confidence intervals for estimators based on the main dataset only.

There exists a connection between the MAC and MVAR approaches. When no heterogeneity is assumed ($\tau=0$ and $\delta=0$) and $\theta = \psi$, both methods lead to the linear combination among $\sum_{i=0}^D\hat\theta_d$ with the smallest variance, where $\hat\theta_0$ is obtained from the main ($d=0$) and $\hat\theta_0$ are obtained form external datasets ($d=1,\ldots,D$). This is also equivalent to pooling normal observations into a single sample.

This paper has focused mainly on statistical methods. However, more aspects should be considered when choosing a strategy for coping with unplanned trial disruptions using auxiliary data. Are we dealing with an exploratory or a confirmatory trial? What importance do we attribute to external credibility, perhaps towards regulatory or reimbursement authorities, as opposed to purposes that are purely internal to the institution running the trial? The very nature of {\it unplanned} trial disruptions seems to preclude pre-specification of methods, which is one of the most powerful principles for avoiding bias and establishing external credibility. However, for blinded trials, mitigation strategies can still be put in place before unblinding. For newly planned trials, pharmaceutical trial sponsors have started to take into account putative intercurrent events at a globally disruptive scale, in the definition of the trial’s estimand; see \cite{ich, VanLancker2022Estimands}.

There are variations to the theme. For example, for trials comparing a new intervention with a control, no external information may be available on the new intervention – but much on the control. The methods presented in this paper could then be employed for the control arm. This connects the situation of unplanned trial disruptions with the theme of external controls, for which a rich literature exists (see \cite{burger2021extcontr} for a recent discussion). Another aspect that we have only barely touched upon is the option to elicit expert opinions to retrieve additional information. We refer to \cite{garthwaite2005statistical, o2006uncertain} and note that this approach may be useful in some situations, but it will not be acceptable in confirmatory situations of regulatory relevance. An interesting question is which of these approaches could or should be considered a priori regardless of any trial disruptions, simply to make a trial more efficient by using the maximum of information available. In fact, all the approaches discussed in this paper had been originally developed and motivated by this uninterrupted situation. It will be interesting to see whether the global COVID-19 pandemic, as catastrophic as it is, can enhance the adoption of innovative methodology for making inferences in clinical trials. 

\newpage
\subsection*{Acknowledgements}
\begin{wrapfigure}{r}{.2\textwidth}
  \begin{center}
    \includegraphics[width=0.2\textwidth]{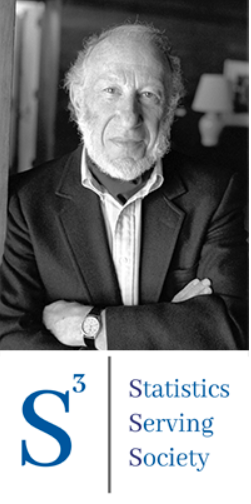}
  \end{center}
\end{wrapfigure}
The authors thank the National Institute of Statistical Sciences for facilitating this work on Coping with Information Loss and the Use of Auxiliary Sources of Data, which is part of the Ingram Olkin Forum Series on Unplanned Clinical Trial Disruptions. The authors would also like to recognize the organizers of this forum series (those who are not an author on this paper): Chris Jennison and Adam Lane as well as the speaker at the motivating workshop who was not an author on this paper: Heng Li.

K.V.L. is supported by Fulbright Belgium, Belgian American Educational Foundation
and VLAIO (Flemish Innovation and Entrepreneurship) under the Baekeland grant agreement
HBC.2017.0219. S.T. is partially supported by the Department of Health and Human Services of the National Institutes of Health under award number R40MC41748.

\bibliographystyle{chicago}
\bibliography{auxiliary_info_bib}

\begin{thebibliography}{}

\bibitem[\protect\citeauthoryear{Akacha, Branson, Bretz, Dharan, Gallo,
  Gathmann, Hemmings, Jones, Xi, and Zuber}{Akacha et~al.}{2020}]{akacha2020}
Akacha, M., J.~Branson, F.~Bretz, B.~Dharan, P.~Gallo, I.~Gathmann,
  R.~Hemmings, J.~Jones, D.~Xi, and E.~Zuber (2020).
\newblock Challenges in assessing the impact of the {COVID-19} pandemic on the
  integrity and interpretability of clinical trials.
\newblock {\em Statistics in Biopharmaceutical Research\/}~{\em 12\/}(4),
  419--426.

\bibitem[\protect\citeauthoryear{Albertus}{Albertus}{2022}]{Albertus2022}
Albertus, M. (2022).
\newblock Asymptotic z and chi-squared tests with auxiliary information.
\newblock {\em Metrika\/}~{\em XX}, xx--xx.

\bibitem[\protect\citeauthoryear{Austin and Mamdani}{Austin and
  Mamdani}{2006}]{austin2006comparison}
Austin, P.~C. and M.~M. Mamdani (2006).
\newblock A comparison of propensity score methods: a case-study estimating the
  effectiveness of {post-AMI} statin use.
\newblock {\em Statistics in medicine\/}~{\em 25\/}(12), 2084--2106.

\bibitem[\protect\citeauthoryear{Azizi, Zheng, Mosquera, Pilote, and
  El~Emam}{Azizi et~al.}{2021}]{synthdata1}
Azizi, Z., M.~Zheng, L.~Mosquera, L.~Pilote, and K.~El~Emam (2021).
\newblock Can synthetic data be a proxy for real clinical trial data ? a
  validation study.
\newblock {\em BMJ Open\/}~{\em 11\/}(4), e043497.

\bibitem[\protect\citeauthoryear{Bang and Robins}{Bang and
  Robins}{2005}]{bang2005doubly}
Bang, H. and J.~M. Robins (2005).
\newblock Doubly robust estimation in missing data and causal inference models.
\newblock {\em Biometrics\/}~{\em 61\/}(4), 962--973.

\bibitem[\protect\citeauthoryear{Benkeser, D{\'\i}az, Luedtke, Segal,
  Scharfstein, and Rosenblum}{Benkeser et~al.}{2021}]{benkeser2021improving}
Benkeser, D., I.~D{\'\i}az, A.~Luedtke, J.~Segal, D.~Scharfstein, and
  M.~Rosenblum (2021).
\newblock Improving precision and power in randomized trials for covid-19
  treatments using covariate adjustment, for binary, ordinal, and time-to-event
  outcomes.
\newblock {\em Biometrics\/}~{\em 77\/}(4), 1467--1481.

\bibitem[\protect\citeauthoryear{Burger, Gerlinger, Harbron, Koch, Posch,
  Rochon, and Schiel}{Burger et~al.}{2021}]{burger2021extcontr}
Burger, H.~U., C.~Gerlinger, C.~Harbron, A.~Koch, M.~Posch, J.~Rochon, and
  A.~Schiel (2021).
\newblock The use of external controls: To what extent can it currently be
  recommended?
\newblock {\em Pharmaceutical Statistics\/}~{\em 20\/}(6), 1002--1016.

\bibitem[\protect\citeauthoryear{Chen and Ibrahim}{Chen and
  Ibrahim}{2006}]{chen2006}
Chen, M.-H. and J.~G. Ibrahim (2006).
\newblock {The relationship between the power prior and hierarchical models}.
\newblock {\em Bayesian Analysis\/}~{\em 1\/}(3), 551 -- 574.

\bibitem[\protect\citeauthoryear{Dmitriev, Tarassenko, and Ustinov}{Dmitriev
  et~al.}{2014}]{Dmitriev2014}
Dmitriev, Y., P.~Tarassenko, and Y.~Ustinov (2014).
\newblock On estimation of linear functional by utilizing a prior guess.
\newblock In A.~Dudin, A.~Nazarov, R.~Yakupov, and A.~Gortsev (Eds.), {\em
  Information Technologies and Mathematical Modelling}, Cham, pp.\  82--90.
  Springer International Publishing.

\bibitem[\protect\citeauthoryear{Duan, Ye, and Smith}{Duan
  et~al.}{2006}]{duan2006}
Duan, Y., K.~Ye, and E.~P. Smith (2006).
\newblock Evaluating water quality using power priors to incorporate historical
  information.
\newblock {\em Environmetrics\/}~{\em 17\/}(1), 95--106.

\bibitem[\protect\citeauthoryear{Engel and Walstra}{Engel and
  Walstra}{1991}]{engel1991}
Engel, B. and P.~Walstra (1991).
\newblock Increasing precision or reducing expense in regression experiments by
  using information from a concomitant variable.
\newblock {\em Biometrics\/}~{\em 47}, 13--20.

\bibitem[\protect\citeauthoryear{Epstein, Schechtman, Kilanowski, Ramel,
  Moursi, Quattrin, Cook, Eneli, Pratt, Geller, et~al.}{Epstein
  et~al.}{2021}]{EPSTEIN2021106497}
Epstein, L.~H., K.~B. Schechtman, C.~Kilanowski, M.~Ramel, N.~A. Moursi,
  T.~Quattrin, S.~R. Cook, I.~U. Eneli, C.~Pratt, N.~Geller, et~al. (2021).
\newblock Implementing family-based behavioral treatment in the pediatric
  primary care setting: Design of the {PLAN} study.
\newblock {\em Contemporary clinical trials\/}~{\em 109}, 106497.

\bibitem[\protect\citeauthoryear{Fogel}{Fogel}{2018}]{Fogel2018}
Fogel, D.~B. (2018, Sep).
\newblock {{F}actors associated with clinical trials that fail and
  opportunities for improving the likelihood of success: {A} review}.
\newblock {\em Contemp Clin Trials Commun\/}~{\em 11}, 156--164.

\bibitem[\protect\citeauthoryear{{Food and Drug Administration}}{{Food and Drug
  Administration}}{2010}]{fda}
{Food and Drug Administration} (2010).
\newblock {Guidance for Industry and FDA Staff: Guidance for the Use of
  Bayesian Statistics in Medical Device Clinical Trials}.
\newblock
  \url{https://www.fda.gov/regulatory-information/search-fda-guidance-documents/guidance-use-bayesian-statistics-medical-device-clinical-trials}.
\newblock Accessed: 10.12.2021.

\bibitem[\protect\citeauthoryear{Galbraith and Marschner}{Galbraith and
  Marschner}{2003}]{galbraith2003}
Galbraith, S. and I.~C. Marschner (2003).
\newblock Interim analysis of continuous long-term endpoints in clinical trials
  with longitudinal outcomes.
\newblock {\em Statistics in Medicine\/}~{\em 22\/}(11), 1787--1805.

\bibitem[\protect\citeauthoryear{Garthwaite, Kadane, and O'Hagan}{Garthwaite
  et~al.}{2005}]{garthwaite2005statistical}
Garthwaite, P.~H., J.~B. Kadane, and A.~O'Hagan (2005).
\newblock Statistical methods for eliciting probability distributions.
\newblock {\em Journal of the American Statistical Association\/}~{\em
  100\/}(470), 680--701.

\bibitem[\protect\citeauthoryear{Gravestock, Held, and behalf of~the
  COMBACTE-Net~consortium}{Gravestock et~al.}{2017}]{gravestock2017}
Gravestock, I., L.~Held, and O.~behalf of~the COMBACTE-Net~consortium (2017).
\newblock Adaptive power priors with empirical {Bayes} for clinical trials.
\newblock {\em Pharmaceutical Statistics\/}~{\em 16\/}(5), 349--360.

\bibitem[\protect\citeauthoryear{Hawila and Berg}{Hawila and
  Berg}{2021}]{Hawila21}
Hawila, N. and A.~Berg (2021, 05).
\newblock {{A}ssessing the impact of {C}{O}{V}{I}{D}-19 on registered
  interventional clinical trials}.
\newblock {\em Clin Transl Sci\/}~{\em 14\/}(3), 1147--1154.

\bibitem[\protect\citeauthoryear{Higgins, Thomas, Chandler, Cumpston, Li, Page,
  and Welch}{Higgins et~al.}{2021}]{cochrane}
Higgins, J.~P., J.~Thomas, J.~Chandler, M.~Cumpston, T.~Li, M.~J. Page, and
  V.~A. Welch (2021).
\newblock Cochrane handbook for systematic reviews of interventions version 6.2
  (updated february 2021).
\newblock \url{https://www.training.cochrane.org/handbook}.
\newblock Accessed: 28.02.2022.

\bibitem[\protect\citeauthoryear{Hobbs, Carlin, Mandrekar, and Sargent}{Hobbs
  et~al.}{2011}]{hobbs2011}
Hobbs, B.~P., B.~P. Carlin, S.~J. Mandrekar, and D.~J. Sargent (2011).
\newblock {Hierarchical Commensurate and Power Prior Models for Adaptive
  Incorporation of Historical Information in Clinical Trials}.
\newblock {\em Biometrics\/}~{\em 67\/}(3), 1047--1056.

\bibitem[\protect\citeauthoryear{Hobbs, Sargent, and Carlin}{Hobbs
  et~al.}{2012}]{hobbs2012}
Hobbs, B.~P., D.~J. Sargent, and B.~P. Carlin (2012).
\newblock Commensurate priors for incorporating historical information in
  clinical trials using general and generalized linear models.
\newblock {\em Bayesian Analysis\/}~{\em 7\/}(3), 639 -- 674.

\bibitem[\protect\citeauthoryear{Hua, Janocha, Severin, Wei, and
  Vandemeulebroecke}{Hua et~al.}{2022}]{hua2021phase}
Hua, E., R.~Janocha, T.~Severin, J.~Wei, and M.~Vandemeulebroecke (2022).
\newblock {A Phase 3 trial analysis proposal for mitigating the impact of the
  COVID-19 pandemic}.
\newblock {\em Statistics in Biopharmaceutical Research\/}~{\em 14\/}(1),
  80--86.

\bibitem[\protect\citeauthoryear{Ibrahim, Chen, Gwon, and Chen}{Ibrahim
  et~al.}{2015}]{ibrahim2015}
Ibrahim, J., M.~H. Chen, Y.~Gwon, and F.~Chen (2015, 09).
\newblock {The Power Prior: Theory and Applications}.
\newblock {\em Statistics in Medicine\/}~{\em 34\/}(28), 3724--3749.

\bibitem[\protect\citeauthoryear{Ibrahim and Chen}{Ibrahim and
  Chen}{2000}]{ibrahim2000}
Ibrahim, J.~G. and M.-H. Chen (2000).
\newblock {Power Prior Distributions for Regression Models}.
\newblock {\em Statistical Science\/}~{\em 15\/}(1), 46--60.

\bibitem[\protect\citeauthoryear{{International Council for Harmonisation of
  Technical Requirements for Pharmaceuticals for Human Use}}{{International
  Council for Harmonisation of Technical Requirements for Pharmaceuticals for
  Human Use}}{2019}]{ich}
{International Council for Harmonisation of Technical Requirements for
  Pharmaceuticals for Human Use} (2019).
\newblock {ICH Harmonised Guideline E9 (R1): Estimands} and sensitivity
  analysis in clinical trials.
\newblock
  \url{https://database.ich.org/sites/default/files/E9-R1\_Step4\_Guideline\_2019\_1203.pdf}.
\newblock Accessed: 28.02.2022.

\bibitem[\protect\citeauthoryear{James, Harbron, Branson, and Sundler}{James
  et~al.}{2021}]{synthdata2}
James, S., C.~Harbron, J.~Branson, and M.~Sundler (2021).
\newblock Synthetic data use: exploring use cases to optimise data utility.
\newblock {\em Discover Artificial Intelligence\/}~{\em 1\/}(1), 1--13.

\bibitem[\protect\citeauthoryear{Kass and Raftery}{Kass and
  Raftery}{1995}]{kass1995}
Kass, R.~E. and A.~E. Raftery (1995).
\newblock Bayes factors.
\newblock {\em Journal of the American Statistical Association\/}~{\em
  90\/}(430), 773--795.

\bibitem[\protect\citeauthoryear{Kopp-Schneider, Calderazzo, and
  Wiesenfarth}{Kopp-Schneider et~al.}{2020}]{kopp2020}
Kopp-Schneider, A., S.~Calderazzo, and M.~Wiesenfarth (2020).
\newblock {Power gains by using external information in clinical trials are
  typically not possible when requiring strict type I error control}.
\newblock {\em Biometrical Journal\/}~{\em 62\/}(2), 361--374.

\bibitem[\protect\citeauthoryear{Kunz, J{\"o}rgens, Bretz, Stallard,
  Van~Lancker, Xi, Zohar, Gerlinger, and Friede}{Kunz et~al.}{2020}]{kunz2020}
Kunz, C.~U., S.~J{\"o}rgens, F.~Bretz, N.~Stallard, K.~Van~Lancker, D.~Xi,
  S.~Zohar, C.~Gerlinger, and T.~Friede (2020).
\newblock {Clinical trials impacted by the COVID-19 pandemic: Adaptive designs
  to the rescue?}
\newblock {\em Statistics in Biopharmaceutical Research\/}~{\em 12\/}(4),
  461--477.

\bibitem[\protect\citeauthoryear{Ledford}{Ledford}{2020}]{ledford2020}
Ledford, H. (2020).
\newblock Coronavirus shuts down trials of drugs for multiple other diseases.
\newblock {\em Nature\/}~{\em 580\/}(7801), 15--17.

\bibitem[\protect\citeauthoryear{Lunn, Barrett, Sweeting, and Thompson}{Lunn
  et~al.}{2013}]{lunn2013}
Lunn, D., J.~Barrett, M.~Sweeting, and S.~Thompson (2013).
\newblock {Fully Bayesian hierarchical modelling in two stages, with
  application to meta-analysis}.
\newblock {\em Journal of the Royal Statistical Society. Series C, Applied
  statistics\/}~{\em 62\/}(4), 551--572.

\bibitem[\protect\citeauthoryear{Marschner and Becker}{Marschner and
  Becker}{2001}]{marschner2001}
Marschner, I.~C. and S.~L. Becker (2001).
\newblock Interim monitoring of clinical trials based on long-term binary
  endpoints.
\newblock {\em Statistics in Medicine\/}~{\em 20\/}(2), 177--192.

\bibitem[\protect\citeauthoryear{Neuenschwander, Branson, and
  Spiegelhalter}{Neuenschwander et~al.}{2009}]{neuenschwander2009}
Neuenschwander, B., M.~Branson, and D.~J. Spiegelhalter (2009).
\newblock A note on the power prior.
\newblock {\em Statistics in Medicine\/}~{\em 28\/}(28), 3562--3566.

\bibitem[\protect\citeauthoryear{Neuenschwander, Capkun-Niggli, Branson, and
  Spiegelhalter}{Neuenschwander et~al.}{2010}]{neuenschwander2010}
Neuenschwander, B., G.~Capkun-Niggli, M.~Branson, and D.~J. Spiegelhalter
  (2010).
\newblock Summarizing historical information on controls in clinical trials.
\newblock {\em Clinical Trials\/}~{\em 7\/}(1), 5--18.
\newblock PMID: 20156954.

\bibitem[\protect\citeauthoryear{Neuenschwander, Roychoudhury, and
  Schmidli}{Neuenschwander et~al.}{2016}]{neuenschwander2016}
Neuenschwander, B., S.~Roychoudhury, and H.~Schmidli (2016).
\newblock On the use of co-data in clinical trials.
\newblock {\em Statistics in Biopharmaceutical Research\/}~{\em 8\/}(3),
  345--354.

\bibitem[\protect\citeauthoryear{Neuenschwander and Schmidli}{Neuenschwander
  and Schmidli}{2020}]{neuenschwander2020}
Neuenschwander, B. and H.~Schmidli (2020).
\newblock Use of historical data.
\newblock In {\em Bayesian Methods in Pharmaceutical Research}, pp.\  111--137.
  Chapman and Hall/CRC.

\bibitem[\protect\citeauthoryear{Nicholas, Straube, Schmidli, Schneider, and
  Friede}{Nicholas et~al.}{2011}]{nicholas2011trends}
Nicholas, R., S.~Straube, H.~Schmidli, S.~Schneider, and T.~Friede (2011).
\newblock Trends in annualized relapse rates in relapsing--remitting multiple
  sclerosis and consequences for clinical trial design.
\newblock {\em Multiple Sclerosis Journal\/}~{\em 17\/}(10), 1211--1217.

\bibitem[\protect\citeauthoryear{O'Hagan, Buck, Daneshkhah, Eiser, Garthwaite,
  Jenkinson, Oakley, and Rakow}{O'Hagan et~al.}{2006}]{o2006uncertain}
O'Hagan, A., C.~E. Buck, A.~Daneshkhah, J.~R. Eiser, P.~H. Garthwaite, D.~J.
  Jenkinson, J.~E. Oakley, and T.~Rakow (2006).
\newblock {\em Uncertain judgements: eliciting experts' probabilities}.
\newblock John Wiley \& Sons.

\bibitem[\protect\citeauthoryear{Ollier, Morita, Ursino, and Zohar}{Ollier
  et~al.}{2020}]{ollier2020}
Ollier, A., S.~Morita, M.~Ursino, and S.~Zohar (2020).
\newblock An adaptive power prior for sequential clinical trials--application
  to bridging studies.
\newblock {\em Statistical methods in medical research\/}~{\em 29\/}(8),
  2282--2294.

\bibitem[\protect\citeauthoryear{Ollier, Zohar, Morita, and Ursino}{Ollier
  et~al.}{2021}]{ollier2021}
Ollier, A., S.~Zohar, S.~Morita, and M.~Ursino (2021).
\newblock Estimating similarity of dose--response relationships in phase {I}
  clinical trials—case study in bridging data package.
\newblock {\em International journal of environmental research and public
  health\/}~{\em 18\/}(4), 1639.

\bibitem[\protect\citeauthoryear{Rosenbaum, Rosenbaum, and Briskman}{Rosenbaum
  et~al.}{2010}]{rosenbaum2010design}
Rosenbaum, P.~R., P.~Rosenbaum, and Briskman (2010).
\newblock {\em Design of observational studies}, Volume~10.
\newblock Springer.

\bibitem[\protect\citeauthoryear{Rosenblum, Qian, Du, , and Qiu}{Rosenblum
  et~al.}{2015}]{rosenblum2015}
Rosenblum, M., T.~Qian, Y.~Du, , and H.~Qiu (2015).
\newblock Adaptive enrichment designs for randomized trials with delayed
  endpoints, using locally efficient estimators to improve precision.
\newblock Johns hopkins university, department of biostatistics working papers.
  working paper 275 https://biostats.bepress.com/jhubiostat/paper275, Johns
  Hopkins University.

\bibitem[\protect\citeauthoryear{R{\"o}ver, Bender, Dias, Schmid, Schmidli,
  Sturtz, Weber, and Friede}{R{\"o}ver et~al.}{2021}]{rover2021}
R{\"o}ver, C., R.~Bender, S.~Dias, C.~H. Schmid, H.~Schmidli, S.~Sturtz,
  S.~Weber, and T.~Friede (2021).
\newblock {On weakly informative prior distributions for the heterogeneity
  parameter in Bayesian random-effects meta-analysis}.
\newblock {\em Research Synthesis Methods\/}~{\em 12\/}(4), 448--474.

\bibitem[\protect\citeauthoryear{Schmidli, Gsteiger, Roychoudhury, O'Hagan,
  Spiegelhalter, and Neuenschwander}{Schmidli et~al.}{2014}]{schmidli2014}
Schmidli, H., S.~Gsteiger, S.~Roychoudhury, A.~O'Hagan, D.~Spiegelhalter, and
  B.~Neuenschwander (2014).
\newblock Robust meta-analytic-predictive priors in clinical trials with
  historical control information.
\newblock {\em Biometrics\/}~{\em 70\/}(4), 1023--1032.

\bibitem[\protect\citeauthoryear{Stallard}{Stallard}{2010}]{stallard2010}
Stallard, N. (2010).
\newblock A confirmatory seamless phase {II/III} clinical trial design
  incorporating short-term endpoint information.
\newblock {\em Statistics in Medicine\/}~{\em 29\/}(9), 959--971.

\bibitem[\protect\citeauthoryear{Takeda and Morita}{Takeda and
  Morita}{2015}]{takeda2015}
Takeda, K. and S.~Morita (2015).
\newblock Incorporating historical data in {Bayesian} phase {I} trial design:
  The caucasian-to-asian toxicity tolerability problem.
\newblock {\em Therapeutic Innovation \& Regulatory Science\/}~{\em 49\/}(1),
  93--99.

\bibitem[\protect\citeauthoryear{Tarima and Pavlov}{Tarima and
  Pavlov}{2006}]{Tarima2006}
Tarima, S. and D.~Pavlov (2006).
\newblock Using auxiliary information in statistical function estimation.
\newblock {\em ESAIM: Probab. Stat.\/}~{\em 10}, 11--23.

\bibitem[\protect\citeauthoryear{Tarima, Slavova, Fritsch, and Hall}{Tarima
  et~al.}{2007}]{Tarima2007}
Tarima, S., S.~Slavova, T.~Fritsch, and L.~Hall (2007).
\newblock Probability estimation when some observations are grouped.
\newblock {\em Stat. Med.\/}~{\em 26\/}(8), 1745--1761.

\bibitem[\protect\citeauthoryear{Tarima, Tuyishimire, Sparapani, Rein, and
  Meurer}{Tarima et~al.}{2020}]{tarima2020estimation}
Tarima, S., B.~Tuyishimire, R.~Sparapani, L.~Rein, and J.~Meurer (2020).
\newblock Estimation combining unbiased and possibly biased estimators.
\newblock {\em Journal of Statistical Theory and Practice\/}~{\em 14\/}(2),
  1--20.

\bibitem[\protect\citeauthoryear{van Dorn}{van Dorn}{2020}]{vanDorn2020}
van Dorn, A. (2020, 08).
\newblock {{C}{O}{V}{I}{D}-19 and readjusting clinical trials}.
\newblock {\em Lancet\/}~{\em 396\/}(10250), 523--524.

\bibitem[\protect\citeauthoryear{Van~Lancker, Tarima, Bartlett, Bauer,
  Bharani-Dharan, Bretz, Flournoy, Michiels, Parra, Rosenberger, and
  Cro}{Van~Lancker et~al.}{2022}]{VanLancker2022Estimands}
Van~Lancker, K., S.~Tarima, J.~Bartlett, M.~Bauer, B.~Bharani-Dharan, F.~Bretz,
  N.~Flournoy, H.~Michiels, C.~O. Parra, J.~L. Rosenberger, and S.~Cro (2022).
\newblock Estimands and their estimators for clinical trials impacted by the
  covid-19 pandemic: A report from the niss ingram olkin forum series on
  unplanned clinical trial disruptions.

\bibitem[\protect\citeauthoryear{Van~Lancker, Vandebosch, and
  Vansteelandt}{Van~Lancker et~al.}{2020}]{vanlancker2020}
Van~Lancker, K., A.~Vandebosch, and S.~Vansteelandt (2020).
\newblock Improving interim decisions in randomized trials by exploiting
  information on short-term endpoints and prognostic baseline covariates.
\newblock {\em Pharmaceutical Statistics\/}~{\em 19\/}(5), 583--601.

\bibitem[\protect\citeauthoryear{Viele, Berry, Neuenschwander, Amzal, Chen,
  Enas, Hobbs, Ibrahim, Kinnersley, Lindborg, Micallef, Roychoudhury, and
  Thompson}{Viele et~al.}{2014}]{viele2014}
Viele, K., S.~Berry, B.~Neuenschwander, B.~Amzal, F.~Chen, N.~Enas, B.~Hobbs,
  J.~G. Ibrahim, N.~Kinnersley, S.~Lindborg, S.~Micallef, S.~Roychoudhury, and
  L.~Thompson (2014).
\newblock Use of historical control data for assessing treatment effects in
  clinical trials.
\newblock {\em Pharmaceutical Statistics\/}~{\em 13\/}(1), 41--54.

\bibitem[\protect\citeauthoryear{Zenkova and Krainova}{Zenkova and
  Krainova}{2017}]{Zenkova2017}
Zenkova, Z. and E.~Krainova (2017).
\newblock Estimating the net premium using additional information about a
  quantile of the cumulative distribution function.
\newblock {\em Bus. Inform.\/}~{\em 42\/}(4), 55--63.

\end{thebibliography}

\appendix

\section{Software}

The illustrative example was analysed with R software (\hyperlink{R}{http://www.r-project.org})  v. 4.1.1. R Code is available at \url{github.com/reidcw/NISS-Information-Loss}.

\section{Appendix}
\subsection{Estimates and intervals on pBMI scale}
\begin{figure}[ht]
    \includegraphics[width=1\linewidth]{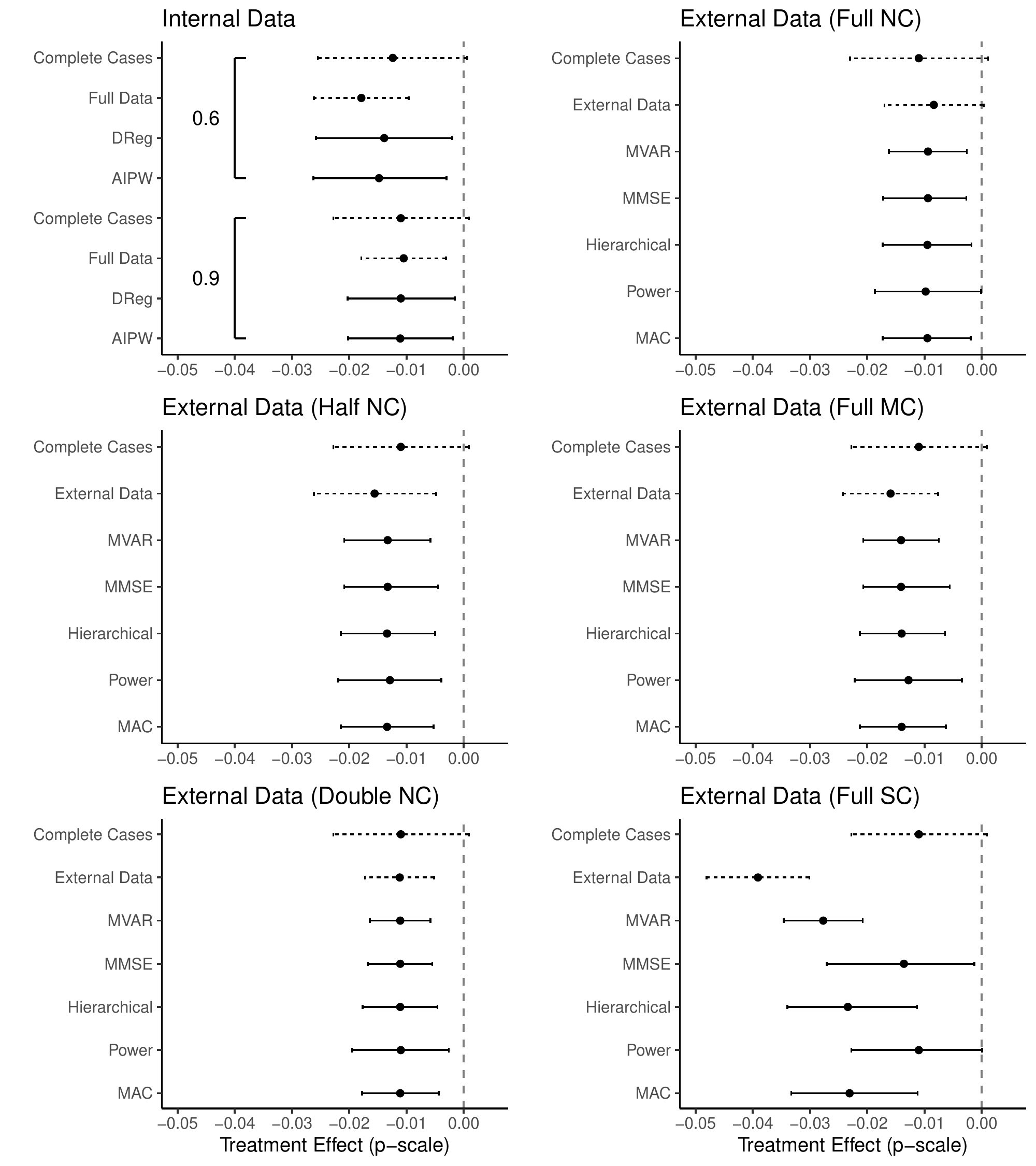}
    \caption{Analysis of the pBMI values: Estimates and 95\% intervals for the treatment effect parameter at 24 months. Confidence intervals are shown for frequentist methods, while symmetric credible intervals are shown for Bayesian methods. In the case of methods leveraging external information, results with incorporation of external data as well as using only the complete cases are reported. For the internal methods, the result based on the full data-set prior to setting any value to missing (Full data) is included for illustrative purposes. The correlation between zBMI2 and zBMI3 (from which pBMI2 and pBMI3 are obtained) is equal to 0.9 unless otherwise specified.}
    \label{fig:resplots_p}
\end{figure}

Supplementary Figure \ref{fig:resplots_p} reports 95\% confidence intervals (on pediatric BMI percentile scale) of using internal and external auxiliary information, respectively.

\end{document}